\documentclass{aa501}

\usepackage{psfig}

\begin{document}
%
\def\lfir{$L_{\rm FIR}$}
\def\mabs{M$_{\rm abs}$}
\def\etal{et al.}
\def\cbeta{$c_{\rm H\beta}$}
\def\av{A$_{\rm v}$}
\def\flam{$F_{\lambda}$}
\def\ilam{$I_{\lambda}$}
\def\teff{\ifmmode T_{\rm eff} \else $T_{\mathrm{eff}}$\fi}
\def\lg{$\log g$}
\def\feh{$\mathrm{[Fe/H]}$}
\def\mh{$\mathrm{[M/H]}$}
\def\ltsima{$\buildrel<\over\sim$}
\def\lsim{\lower.5ex\hbox{\ltsima}}

\newcommand{\hii}{H~{\sc ii}}
\newcommand{\ha}{\ifmmode {\rm H}\alpha \else H$\alpha$\fi}
\newcommand{\hb}{\ifmmode {\rm H}\beta \else H$\beta$\fi}
\newcommand{\whb}{\ifmmode W({\rm H}\beta) \else $W({\rm H}\beta)$\fi}
\newcommand{\wwr}{\ifmmode W({\rm WR}) \else $W({\rm WR})$\fi}
\newcommand{\iwr}{\ifmmode I({\rm WR})/I(\hb) \else $I({\rm WR})/I(\hb)$\fi}
\newcommand{\lya}{Lyman-$\alpha$}
\newcommand{\hei}{He~{\sc i}}
\newcommand{\Hei}{He~{\sc i} $\lambda$4471}
\newcommand{\heii}{He~{\sc ii}}
\newcommand{\Heii}{He~{\sc ii} $\lambda$4686}
\newcommand{\qh}{\ifmmode q({\rm H}) \else $q({\rm H})$\fi}
\newcommand{\qhe}{\ifmmode q({\rm He^0}) \else $q({\rm He^0})$\fi}
\newcommand{\qhep}{\ifmmode q({\rm He^+}) \else $q({\rm He^+})$\fi}
\newcommand{\Qh}{\ifmmode Q({\rm H}) \else $Q({\rm H})$\fi}
\newcommand{\Qhe}{\ifmmode Q({\rm He^0}) \else $Q({\rm He^0})$\fi}
\newcommand{\Qhep}{\ifmmode Q({\rm He^+}) \else $Q({\rm He^+})$\fi}
\newcommand{\Qhtwo}{\ifmmode Q({\rm LW}) \else $Q({\rm LW})$\fi}
\newcommand{\qrathe}{\ifmmode q({\rm He^0})/q({\rm H}) \else $q({\rm He^0})/q({\rm H})$\fi}
\newcommand{\qrathep}{\ifmmode q({\rm He^+})/q({\rm H}) \else $q({\rm He^+})/q({\rm H})$\fi}
\newcommand{\Qrathe}{\ifmmode Q({\rm He^0})/Q({\rm H}) \else $Q({\rm He^0})/Q({\rm H})$\fi}
\newcommand{\Qrathep}{\ifmmode Q({\rm He^+})/Q({\rm H}) \else $Q({\rm He^+})/Q({\rm H})$\fi}
\newcommand{\Qhave}{\ifmmode \bar{Q}({\rm H}) \else $\bar{Q}({\rm H})$\fi}
\newcommand{\Qheave}{\ifmmode \bar{Q}({\rm He^0}) \else $\bar{Q}({\rm He^0})$\fi}
\newcommand{\Qhepave}{\ifmmode \bar{Q}({\rm He^+}) \else $\bar{Q}({\rm He^+})$\fi}
\newcommand{\Qhtwoave}{\ifmmode \bar{Q}({\rm H}_2) \else $\bar{Q}({\rm H}_2)$\fi}
\newcommand{\Qratheave}{\ifmmode \bar{Q}({\rm He^0})/\bar{Q}({\rm H}) \else $\bar{Q}({\rm He^0})/\bar{Q}({\rm H})$\fi}
\newcommand{\Qrathepave}{\ifmmode \bar{Q}({\rm He^+})/\bar{Q}({\rm H}) \else $\bar{Q}({\rm He^+})/\bar{Q}({\rm H})$\fi}

\def\micron{$\mu$m}
\def\kms{km s$^{-1}$}
\def\kmsmpc{km s$^{-1}$ Mpc$^{-1}$}
\def\cmc{cm$^{-3}$}
\def\erg{ergs s$^{-1}$ cm$^{-2}$ \AA$^{-1}$}
\def\ergs{ergs s$^{-1}$}
\def\ergscm{ergs s$^{-1}$ cm$^{-2}$}
\def\msun{\ifmmode M_{\odot} \else M$_{\odot}$\fi}
\def\zsun{\ifmmode Z_{\odot} \else Z$_{\odot}$\fi}
\def\lsun{\ifmmode L_{\odot} \else L$_{\odot}$\fi}

\def\mup{\ifmmode M_{\rm up} \else M$_{\rm up}$\fi}
\def\mlow{\ifmmode M_{\rm low} \else M$_{\rm low}$\fi}
\newcommand{\dt}{\ifmmode \Delta t \else $\Delta t$\fi}
\def\ubvetc{(UBV)$_J$\-(RI)$_C$\- JHKLL$^\prime$M}
\def\basel {{\it B}a{\it S}e{\it L}}
\def\was{CMT$_1$T$_2$}
%
\def\aap{A\&A}
\def\aaps{A\&AS}
\def\aas{A\&AS}
\def\aj{AJ}
\def\apj{ApJ}
\def\apjl{ApJ}
\def\apjs{ApJS}
\def\mnras{MNRAS}
\def\pasp{PASP}
%
\newcommand{\oh}{\ifmmode 12 + \log({\rm O/H}) \else$12 + \log({\rm
O/H})$\fi}
\newcommand{\nii}{[N~{\sc ii}]}
\newcommand{\oi}{[O~{\sc i}]}
\newcommand{\oii}{[O~{\sc ii}]}
\newcommand{\oiii}{[O~{\sc iii}]}
\def\Nii{[N\small II]\normalsize $\lambda\lambda$6548,6584}
\def\Sii{[S~{\sc ii}] $\lambda\lambda$6717,6731}
\def\Siii{[S~{\sc iii}] $\lambda\lambda$9069,9532}
\def\Oii{[O~{\sc ii}] $\lambda$3727}
\def\Oiii{[O~{\sc iii}] $\lambda\lambda$4959,5007}
\newcommand{\Niiib}{N~{\sc iii} $\lambda$4512}
\newcommand{\Nv}{N~{\sc v} $\lambda$4612}
\newcommand{\Niii}{N~{\sc iii} $\lambda$4640}
\newcommand{\Civb}{C~{\sc iv} $\lambda$4658}
\newcommand{\Ciii}{C~{\sc iii} $\lambda$5696}
\newcommand{\Civ}{C~{\sc iv} $\lambda$5808}



\title{VLT observations of metal-rich extra galactic HII regions. 
I. Massive star populations and the upper end of the IMF
\thanks{Based on observations ESO/VLT service observations (65.N-0308
  and 67.B-0197)
}
}

\author{Maximilien Pindao\inst{1} 
  \and Daniel Schaerer \inst{2}
  \and Rosa M.\ Gonz\'alez Delgado \inst{3}
  \and Gra\.{z}yna Stasi\'nska \inst{4}
}

\offprints{D. Schaerer, schaerer@ast.obs-mip.fr}

 \institute{
Observatoire de Gen\`eve, Ch. des Maillettes 51, CH-1290 Sauverny, Switzerland
\and 
Observatoire Midi-Pyr\'en\'ees, Laboratoire d'Astrophysique, UMR
 5572, 14, Av.  E. Belin, F-31400 Toulouse, France 
\and
Instituto de Astrof\'\i sica de Andaluc\'\i a (CSIC), Apdo. 3004, E-18080, Granada, Spain
\and 
LUTH, Observatoire de Meudon, 5, Place Jules Jansses, F-92150 Meudon, France}

\date{Received 20 june 2002/ Accepted 12 august 2002}

\titlerunning{Metal-rich \hii\ regions and the IMF}

\abstract{We have obtained high quality 
FORS1/VLT optical spectra of 85 disk \hii\ regions in the 
nearby spiral galaxies NGC 3351, NGC 3521, NGC 4254, NGC 4303, and NGC 4321.
Our sample of metal-rich \hii\ regions with metallicities close to 
solar and higher reveal the presence of Wolf-Rayet (WR) stars in 27 objects 
from the blue WR bump ($\sim$ 4680 \AA) and 15 additional candidate WR regions.
This provides for the first time a large set of metal-rich WR regions.
\\
Approximately half (14) of the WR regions also show broad \Civ\ emission
attributed to WR stars of the WC subtype.
The simultaneous detection of \Ciii\ emission in 8 of them allows us to
determine an average late WC subtype compatible
with expectations for high metallicities.
Combined with literature data, the metallicity trends of WR features and the WC/WN 
number ratio are discussed.
\\
The WR regions show quite clear trends between their observed WR features and the 
\hb\ emission line. Detailed synthesis models are presented to understand/interpret
these observations. 
In contrast with earlier studies of low metallicity WR galaxies, 
both \wwr\ and \iwr\ are here found to be smaller than ``standard'' predictions from
appropriate evolutionary synthesis models at corresponding metallicities.
Various possibilities 
which could explain this discrepancy are discussed. 
The most likely solution is found with an improved prescription
to predict the line emission from WN stars in synthesis models.
\\
The availability of a fairly large sample of metal-rich WR regions
allows us to improve existing estimates of the upper mass cut-off of the IMF
in a robust way and independently of detailed modeling:
from the observed maximum \hb\ equivalent width of the WR regions we derive 
a {\bf lower limit for \mup\ of 60--90 \msun} in the case of a Salpeter
slope and larger values for steeper IMF slopes.
This constitutes a lower limit on \mup\ as all observational effects known to 
affect potentially the \hb\ equivalent width 
can only reduce the observed \whb. 
\\
From our direct probe of the massive star content we conclude that 
there is at present no evidence for systematic variations of the upper mass 
cut-off of the IMF in metal-rich environments, in contrast to some claims
based on indirect nebular diagnostics.
}

\maketitle

\section{Introduction}
\label{s_intro}

Wolf-Rayet stars (WR) are the descendants of the most massive stars.
Although they live during a short time (Maeder \& Conti 1994) 
these stars have been detected in young stellar systems, such as extragalactic HII regions
(Kunth \& Schild 1986) and the so-called WR galaxies (Conti 1991, 
Schaerer \etal\ 1999b). They
are recognized by the presence of broad stellar emission lines at optical 
wavelengths, mainly at 4680 \AA\ (known as the blue WR bump) and at 5808 \AA\ 
(red WR bump). The blue bump is a blend of N~{\sc v} $\lambda\lambda$4604,4620,
N~{\sc iii} $\lambda\lambda$4634,4641, C~{\sc iii/iv} $\lambda\lambda$4650,4658 and 
\Heii\ lines, that are produced in WR stars of the nitrogen (WN)
and carbon (WC) sequences. In contrast, the red bump is formed only by 
\Civ\ and it is mainly produced by WC stars. The detection of these
features in the integrated spectrum of a stellar system provides a powerful tool 
to date the onset of the burst, and it constitutes the best direct measure of the
upper end of the initial mass function (IMF). Thus, if WR features are found
in the spectra of star forming systems, stars more massive than $M_{\rm WR}$,
where $M_{\rm WR} \sim$ 25 \msun\ for solar metallicity,
must be formed in the burst.

The IMF is one of the fundamental ingredients 
for studies of stellar populations, which has an important bearing on many 
astrophysical studies ranging from cosmology to the understanding of the local
Universe. In particular the value of the IMF slope and the upper mass cut-off 
(\mup) strongly influences the mechanical, radiative, and chemical feedback
from massive stars to the ISM such as the UV light, the ionizing radiation field,
and the production of heavy elements.

A picture of a universal IMF has emerged from numerous works performed in 
the last few years (e.g.\ Gilmore \& Howell 1998 and references therein). Indeed, these 
studies derive
a slope of the IMF close to the Salpeter value for a mass range between 
5 and 60 \msun. This result seems to hold for a variety of objects
and metallicities from very metal poor  up to the solar metallicity,
with the possible exception of a steeper field IMF (Massey \etal\ 1995, 
Tremonti \etal\ 2002).
However,
the IMF in high metallicity (12+log (O/H) $\ga$ (O/H)$_\odot \approx$ 8.92) 
systems is much less well constrained. 
Different indirect methods to derive the slope and \mup\ give contradictory results.

The detection of strong wind resonance UV lines in the integrated spectrum 
of high metallicity nuclear starbursts clearly indicate the formation of massive stars
(Leitherer 1998; Schaerer 2000; Gonz\'alez Delgado 2001). In contrast, the analysis of the 
nebular optical and infrared lines of IR-luminous galaxies and high metallicity \hii\ regions
indicates a softness of the ionizing radiation field that has beeninterpreted as due 
to the lack of stars more massive than $\sim$ 30 \msun\ (Goldader \etal\ 1997; Bresolin
\etal\ 1999; Thornley \etal\ 2000; Coziol \etal\ 2001).    
However, the interpretation of these indirect probes relies strongly on a combination 
of models for stellar atmospheres and interiors, evolutionary synthesis,
and photoionisation, each with several potential shortcomings/difficulties
(cf.\ Garc\'\i a-Vargas 1996, Schaerer 2000, Stasi\'nska 2002).
For example, recently Gonz\'alez Delgado \etal\ (2002) have shown that the above conclusion 
could be an artifact of the failure of WR stellar atmospheres models to correctly predict the 
ionizing radiation field of high metallicity starbursts (see also Castellanos 2001,
Castellanos \etal\ 2002b).

A more direct investigation of the stellar content of metal-rich 
nuclear starbursts has been performed
by Schaerer \etal\ (2000, hereafter SGIT00), using the detection of WR features
to constrain \mup. They found that the observational 
data are compatible with a Salpeter IMF extending to masses \mup\ $\ga$ 40 \msun.  
Most recently, a similar conclusion has been obtained by Bresolin \& Kennicutt (2002, hereafter 
BK02) from observations of high-metallicity HII regions in M83, NGC 3351 and NGC 6384.

Here, we present a direct attempt to determine \mup\ based on the detection of WR features
in metal-rich \hii\ regions of a sample of spiral galaxies.
To obtain statistically significant conclusions about \mup\ and the slope of the IMF,
a large sample of \hii\ regions needs to be observed. 
For coeval star formation with a Salpeter IMF and \mup=120 \msun\ at 
metallicities above solar, $\sim$ 60 to 80 \% (depending on the evolutionary scenario 
and age of the region) of the \hii\ regions are expected to exhibit WR signatures 
(Meynet 1995; Schaerer \& Vacca 1998, hereafter SV98).
Thus, to find $\ga$ 40 regions with WR stars (our initial aim)
a sample of at least 5-7 galaxies with $\ga$ 10 \hii\ regions 
per galaxy needs to be observed.  
Spectra of high S/N (at least 30) in the continuum are also required to obtain an accurate
measure of the WR features. For this propose, we have selected the nearby spiral galaxies
NGC 3351, NGC 3521, NGC 4254, NGC 4303 and NGC4321, which have 
have sufficient number of disk \hii\ regions of high-metallicity, as known from 
earlier studies.

Our observations have indeed allowed to find a large number of metal-rich WR \hii\ regions.
The analysis of their massive star content is the main aim of the present paper.
Quite independently of the detailed modeling undertaken below, our sample combined
with additional WR regions from Bresolin \& Kennicutt (2002) allow us to derive
a fairly robust {\em lower limit} on the upper mass cut-off of the IMF in these
metal-rich environments (see Sect.\ \ref{s_imf}).

The structure of the paper is as follows: 
The sample selection, observations and data reduction are described in Sect.\ \ref{s_obs}.
The properties of the \hii\ regions are derived in Sect.\ \ref{s_props}. 
Section \ref{s_wroh} discusses the trends of the WR populations with metallicity. 
Detailed comparisons of the observed WR features with the evolutionary synthesis models are 
presented in Sect.\ \ref{s_models}. 
More model independent constraints on \mup\ are derived in Sect.\ \ref{s_imf}. 
Our main results and conclusions are summarised in Sect.\ \ref{s_conclude}.  
 
\section{Sample selection, observations and reduction}
\label{s_obs}

%
\begin{table*}[htb]
\caption{Galaxy sample}
\label{tab_sample}
\begin{center}
\begin{tabular}{llllllllll} \hline
Galaxy   & NED type and activity &$\alpha$ (J2000) & $\delta$ (J2000) & $v_r$         & distance  \\ 
         &      &                 &                  & [km s$^{-1}$] & [Mpc]     \\ \hline
NGC 3351 & SB(r)b, HII Sbrst   & 10h43m57.8s & +11d42m14s &  778 & 10.0 \\
NGC 3521 & SAB(rs)bc, LINER    & 11h05m48.6s & -00d02m09s &  805 &  7.2 \\  
NGC 4254 & SA(s)c             & 12h18m49.5s & +14d24m59s & 2407 & 16.  \\
NGC 4303 & SAB(rs)bc, HII Sy2  & 12h21m54.9s & +04d28m25s & 1566 & 16.  \\
NGC 4321 & SAB(s)bc, LINER HII & 12h22m54.9s & +15d49m21s & 1571 & 15.21\\    
\hline
\end{tabular}
\end{center}
\end{table*}

\subsection{Selection of the HII regions}
Our target galaxies (see Table \ref{tab_sample}) are selected among 
nearby spiral galaxies
where a sufficient number of disk \hii\ regions of high metallicity
are known from the previous studies of Shields et al.\
(1991), Oey \& Kennicutt (1993), and Zaritsky et al.\ (1994).
Inspection of spectra from the two latter studies kindly made
available to us showed that the vast majority of their spectra are 
not deep enough to allow the detection of WR or other stellar 
signatures in the continuum.

Metallicities \oh\ of all known regions were estimated from the 
published \Oii\ and \Oiii\ intensities using the standard $R_{23}$ 
``strong line'' method and various empirical calibrations.
For the FORS1 multi-object spectroscopic observations described below
\hii\ regions with metallicities above solar 
($\log R_{23} \la$ 0.6) were given first priority.
Secondary criteria taken into account in the choice of the known
\hii\ regions were a large \hb\ equivalent width, and
bright continuum flux at $\sim$ 4650 \AA\ as determined from
inspection of the spectra.
This procedure lead to a first selection of 4 to 7 \hii\ regions per galaxy.
Other regions with lower metallicities and/or lower \hb\ equivalent
widths were retained as secondary targets.

Up to 19 slitlets per exposure can be used for spectroscopy with FORS1.
Our primary targets were first positioned using the R-band images (see below)
and the remaining slitlets
were filled whenever possible with secondary targets. If a slitlet was left
without any of our selected regions, we attempted to target other 
\hii\ regions selected from the \ha\ images of Hodge \& Kennicutt (1983).
For each galaxy a nuclear spectrum, to be reported upon later,
was also obtained.


\subsection{Observations}
R band imaging was obtained with FORS1/VLT in april 2000, and was used to
determine the positions of our targeted regions with sufficient
accuracy. 
Subtracting a local average emission from the host galaxy the
R band magnitudes of our target \hii\ regions were determined;
typical magnitudes of $m_R \sim$ 19--21 are found.

The spectroscopic observations of our sample of \hii\ regions were carried out with
FORS1/VLT in the second 2001 trimester. Table \ref{tab_log} gives informations 
about the exact dates and meteorological conditions during the observations.

\begin{table*}[t]
\caption{\footnotesize Log of the observations with meteorological conditions
and exposure times for both grisms} 
\begin{center}
\begin{tabular}{lccccc} \hline
galaxy          & date                  & weather               &
seeing ["]      & exp.\ time blue [s]& exp.\ time red [s]        \\ \hline 
NGC 3351         & 19.04.2001    & photometric   & 0.8-1.0
        & 1700  & 1700                          \\
NGC 3521         & 25.04.2001    & clear                 &
1.6-2.0         & 1800  & 1800                          \\
NGC 4254         & 23.05.2001    & clear                 &
1.1-1.4         & 900   & 900                           \\
NGC 4303         & 23.05.2001    & clear                 &
0.8-1.1         & 750   & 750                           \\
NGC 4321         & 19.06.2001    & photometric   & 1.3-1.5
        & 1050  & 1050                          \\ \hline 
\end{tabular}
\end{center}
\label{tab_log}
\end{table*}

The spectral range from  3600 \AA\ to 1 $\mu$m was covered with a
``blue'' spectrum from 3600 to 6500 \AA\ with grism
300V+10, and a ``red'' spectrum from 6000 to 10000 \AA\ 
with grism 300I+11. The use of a 1\arcsec\ slit width allowed
to get medium spectral resolution of around 6 \AA\ in the blue and 12 \AA\
in the red. 
Due to the limited slit size, a fraction of the total nebular emission
of the regions may be lost. This effect is acounted for in our 
interpretation of the data (Sect.\ \ref{s_models}).
Unless WR stars follow systematically a different spatial distribution 
than other stars responsible for the continuum emission, a possible
loss of continuum light does not alter our analysis.

Exposure times for each galaxy (see Table \ref{tab_log})
were adapted to obtain in the continuum S/N $\sim$ 30 in the blue, 
(needed for a precise measure of the WR bump) and $\sim$ 10 in the red 
(needed to measure the \Siii\ lines).
Spectrophotometric standard stars data were also acquired.

\subsection{Data reduction and analysis}
Reduction was carried out using the IRAF and MIDAS packages. The first steps
consisted in the usual bias subtraction, flatfield division,
and 2D wavelength calibration.
Flux calibration was done using a standard atmospheric extinction curve 
and spectrophotometric standard stars. 
Given that the spectrophotometric standards were not always obtained during
the night of the observations, we estimate an absolute flux accuracy of 
$\sim$ 10 \%.
In addition, due to the optimisation for a maximum multiplex,
the observations were not taken at parallactic angle, leading to a slight
mismatch between the blue and red spectra.
A quantitative analysis of the effetcs of differential refraction
has not been undertaken here.
As the main diagnostics used in the present paper lie in a limited
wavelength range, and the observations have been taken at small airmass,
this should represent a negligible source of uncertainty.

For each \hii\ region, a background including sky emission and 
underlying emission from the galaxy was extracted from the slitlet
sub-image. 
This procedure was non-trivial as this background spectrum had
in most cases to be determined near the edges of the sub-image,
where the wavelength calibration may slightly deviate from the one
of the \hii\ region.
Special care has been taken for the red
spectra, since the sky emission was often several times brighter than the \hii\
region emission. We thus re-calibrated the background emission spectrum
according to the \hii\ region by comparing the position (and sometimes the
intensity) of the sky emission lines. This time-consuming operation gave very
satisfying results and useable spectra up to 1 $\mu$m for almost all \hii\
regions.
The final 1D spectra were generally extracted with a 4 \arcsec\ wide aperture.

Line intensities and equivalent width were obtained by visually placing a
continuum on both sides of the line and then integrating all over this range.
Errors were estimated by moving the continuum upwards by half the value of the
noise near the line position and re-computing the intensity and equivalent
width.

Where possible the following nebular emission lines were measured:
\oii\ $\lambda$3727, the H Balmer line series including \ha\ to H9,
\Hei, \oiii\ $\lambda$4959,5007, \nii\ $\lambda$5201, 
\hei\ $\lambda$5876, \oi\   $\lambda$6300,  \nii\ $\lambda$6548,6584, 
\hei\ $\lambda$6678, \Sii, \hei\ $\lambda$7065, [Ar~{\sc iii}] $\lambda$7136,
\oii\ $\lambda$7325, and \Siii.
If present, broad emission lines at $\lambda \sim$ 4680 \AA\ 
(referred to subsequently as the (blue) WR bump), \Ciii, and \Civ\
indicative of Wolf-Rayet (WR) stars were also measured.
The spectra were also inspected for the presence of stellar absorption
lines like the Ca~{\sc ii} triplet, the CH G band at $\sim$ 4300 \AA, Mg lines
at $\sim$ 5200 \AA, or TiO bands.

The spectra were deredened using the Whitford \etal\ (1958) extinction law
as parametrised by Izotov \etal\ (1994) assuming an underlying
absorption of $W(\hb)=$ 2 \AA\ and an intrinsinc Balmer decrement ratio
of $I(\ha)/I(\hb)=2.86$.

All detailed results including finding charts, line measurements,
and a detailed analysis of the nebular properties
will be published in a forthcoming paper.


%
\begin{table*}[htb]
\caption{Statistics of WR regions}
\label{tab_wr}
\begin{center}
\begin{tabular}{llll|llllll} \hline
Galaxy   &  \# blue bump & \# \Civ & \# \Ciii & cand.\ blue bump & cand.\ \Civ & cand.\ \Ciii\\ 
         &               &         &          &                  &             &  \\ \hline
NGC 3351 & 2  &    &   & 4 &   & 2  \\
NGC 3521 & 4  &  2 & 1 & 6 & 1 & 1  \\
NGC 4254 & 9  &  8 & 1 &   & 1 & 1  \\
NGC 4303 & 9  &  4 & 3 &   & 3 & 2  \\
NGC 4321 & 3  &    & 3 & 5 & 1 & 1  \\
total    & 27 & 14 & 8 &15 & 6 &10 \\
\hline
\end{tabular}
\end{center}
\end{table*}

\section{Properties of the HII region sample}
\label{s_props}
The properties of our galaxy sample as given by the NED database and 
the adopted distances, are summarised in Table \ref{tab_sample}.
For NGC 3351 and the Virgo cluster member NGC 4321 we adopt the Cepheid 
distances from Freedman \etal\ (2001). 
The other two Virgo galaxies (NGC 4254, NGC 4303) are member of the same
subgroup as NGC 4321 (Boselli, private communication). 
We therefore adopt an identical, approximate distance of 16 Mpc.
The distance of NGC 3521 is taken from Tully's (1998) Nearby Galaxy Catalog.

A total of 121 spectra were extracted from the 95 slitlets.
Nebular emission lines were detected in 88 spectra; 85 correspond to extra-nuclear
regions.

\begin{figure}
\centerline{\psfig{file=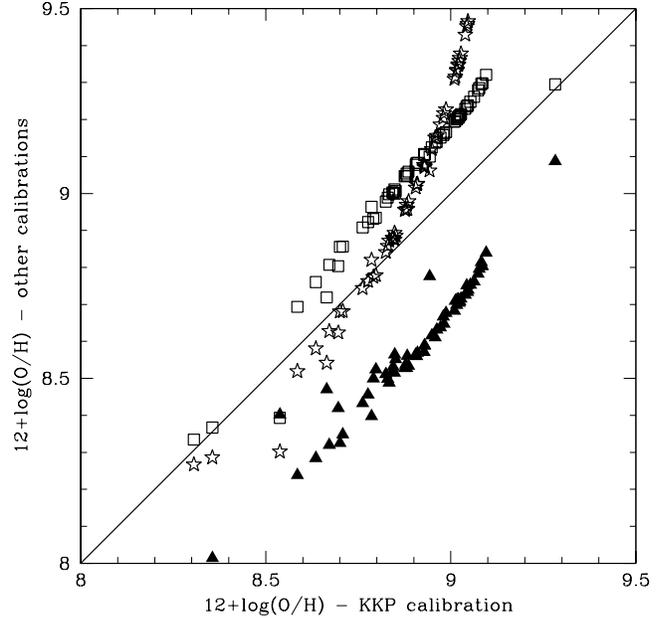,width=8.8cm}}
\caption{Comparison of metallicities O/H of our \hii\ regions derived  
from various empirical calibrations.
The Kobulnicky \etal\ (1999) calibration is taken as a reference (x-axis).
Different symbols show O/H derived from the Pilyugin $P$ method (filled triangles),
and the $R_{23}$ methods of Zaritsky \etal\ (1994, squares),
and Edmunds \& Pagel (1984, stars).
See comments in text.
}
\label{fig_oh_compare}
\end{figure}

\subsection{Metallicities}
The metallicity O/H of the \hii\ regions has been estimated using the following
empirical calibrations:
the calibrations of Kobulnicky \etal\ (1999, hereafter KKP) using \Oiii/\Oii\ 
and (\Oiii+\Oii)/\hb\ (=$R_{23}$) based on the photoionisation model grid of McGaugh (1994), 
the similar $P$-method of Pilyugin (1991), and
the older $R_{23}$ calibrations of Edmunds \& Pagel (1984) and Zaritsky \etal\ (1994).

The O/H abundances obtained from these methods are compared in Fig.\ \ref{fig_oh_compare}.
Unsurprisingly rather large differences are obtained.
As well known, at abundances $\oh \la$ 8.5--8.6 the various $R_{23}$ 
methods yield similar
results, while the differences increase towards higher metallicities 
(see e.g.\ comparison in
Pilyugin 2001). 
Systematically lower values are found from the $P$-method of Pilyugin (2001). 
Although calibrated only for regions with $\oh \la 8.6$, 
this could indicate a systematic overestimate of the absolute metallicities
using the other methods.
To ease comparisons with the recent study of BK02 of metal-rich \hii\ regions
we subsequently adopt the KKP calibration by default except otherwise stated.

\begin{figure}
\centerline{\psfig{file=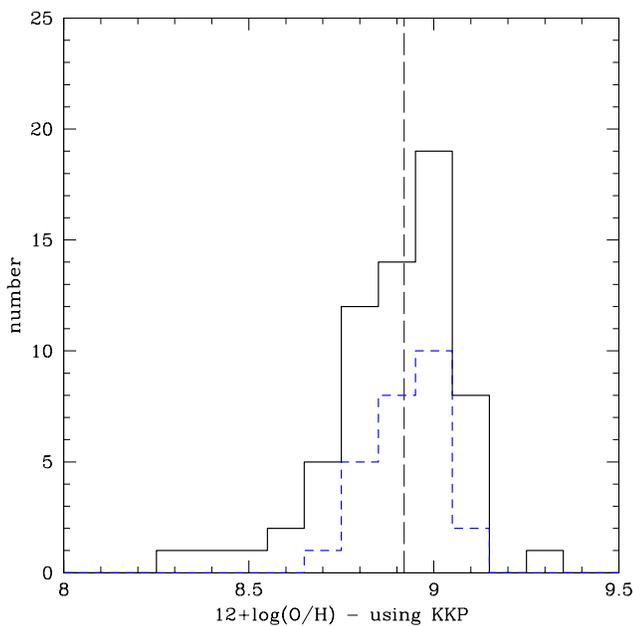,width=8.8cm}}
\caption{Metallicity distribution of our \hii\ region sample 
(solid line) based on the empirical $R_{23}$ calibration of 
Kobulnicky \etal\ (1999).
The distribution of O/H for the WR regions is shown by the dashed line.
The vertical line indicates the solar value adopted in the models
used by these authors.}
\label{fig_oh_histogram}
\end{figure}

The metallicity distribution of our entire sample is shown in Fig.\ \ref{fig_oh_histogram}.
The mean metallicity is $<\oh >= 8.88 \pm 0.22$ (8.57 $\pm$ 0.24) 
using the KKP (Pilyugin's $P$) calibration.
The vertical dashed line indicates the solar value (\oh=8.92) adopted in McGaugh's calculations
used for the calibration of KKP.

\begin{figure}
\centerline{\psfig{file=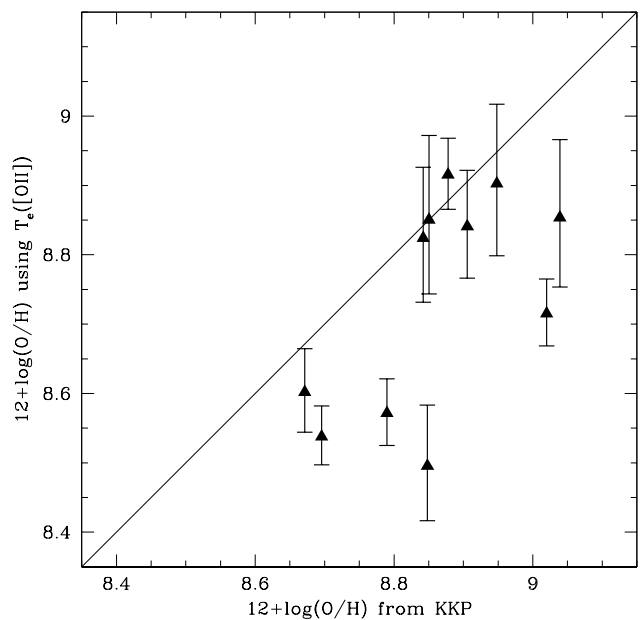,width=8.8cm}}
\caption{Comparison of metallicities O/H derived from the 
empirical calibration of Kobulnicky \etal\ (1999, KKP) with 
the determination in regions with measured electron temperature
$T_e([{\rm OII}])$. 
The errorbars include only the uncertainty on the \oii\ $\lambda$7325/\hb\
ratio.
The diagonal shows the one-to-one relation.
Discussion in text.}
\label{fig_oh_toii}
\end{figure}

\subsubsection{Regions with direct $T_e$ determinations}
The transauroral [O~{\sc ii}] $\lambda$7325 line has been detected 
in 11 \hii\ regions allowing thus a direct determination of the
electron temperature from [O~{\sc ii}] $\lambda$7325/\Oii.
Other potential electron temperature indicators, e.g.\ [S~{\sc iii}] 
$\lambda$6312, [N~{\sc ii}] $\lambda$5755, are too weak or could 
not be measured due to the limited spectral reolution.
Electron densities are determined from \Sii.
$T_e($O~{\sc ii}$)$ and the resulting ionic abundance ratios of 
O$^{++}$/H$^+$ and O$^+$/H$^+$
were derived using this temperature for both ions (the atomic data are those
listed in Stasi\'nska \& Leitherer (1996).

As shown in Fig.\ \ref{fig_oh_toii}
the resulting O/H abundances (assuming O/H $=$ O$^{++}$/H$^+$+O$^+$/H$^+$)
are on average found to be lower than those derived
from the KKP calibration, the largest metallicity being closer to solar.
However, the O/H derived here are lower limits, due
to the strong temperature gradients expected at high metallicities 
(see Stasi\'nska 2002).
A deeper discussion of the abundances in our objects taking into 
account the observational constraints from the entire emission line 
spectrum is deferred to a forthcoming publication.
For the purpose of the present paper, it is sufficient to note that 
the bulk of our \hii\ region sample with low values of $R_{23}$ have
metallicities close to and above solar.

\begin{figure}
\centerline{\psfig{file=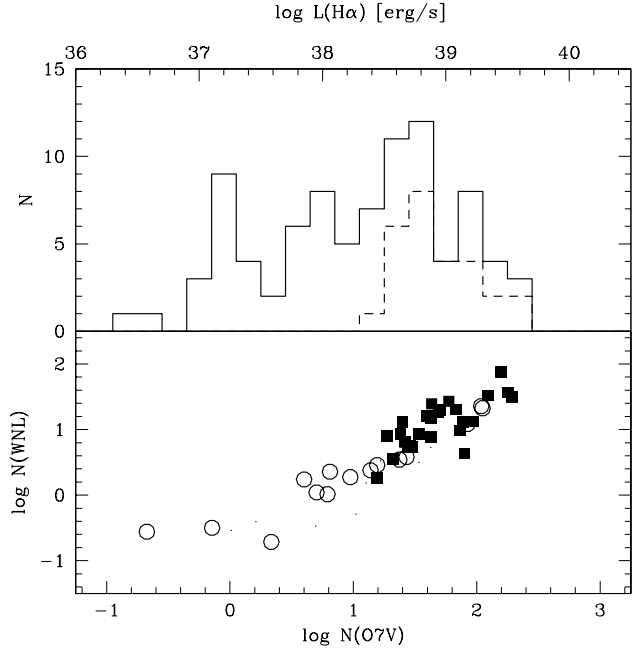,width=8.8cm}}
\caption{{\em Upper panel:} Histogram of \ha\ luminosities of the entire \hii\ region
sample (solid) and the regions with WR detections (dashed).
{\em Lower panel:} Number of WNL stars (derived from blue WR bump assuming WN7 type) 
as a function of the number of equivalent O7V stars derived from \ha\ luminosity.
Note that the x-axis of both plots correspond directly through the definition of N07V 
(see footnote).
Filled (open) symbols indicate regions with certain (candidate) WR detections (see text).
The lower panel shows that secure WR detections are only found for regions with 
$N_{\rm WNL} \protect\ga$ 2--3, as physically expected. 
This indicates that the WR sample in our sample of regions with 
$\log(\ha) \protect\ga 38.2$ is fairly complete (cf.\ text).}
\label{fig_plot_both}
\end{figure}

\subsection{\ha\ luminosities, WR and O star populations}
The histogram of the \ha\ luminosity of the \hii\ regions, as measured from
our spectra, is shown in the upper panel
of Fig.\ \ref{fig_plot_both}.
As seen from this figure, $\sim$ 75 \% of the \hii\ regions correspond to
giant extra-galactic \hii\ regions characterised by $L(\ha) \ga 10^{38}$ erg s$^{-1}$
(Kennicutt 1984, 1991), while the remainder are less luminous objects
similar to normal Galactic \hii\ regions.
The corresponding number of equivalent O7V stars \footnote{
We compute $N_{\rm O7V} = Q_0 / 10^{49.12} = L(\ha) / (1.36\, 10^{-12} \times 10^{49.12}$
taking the Lyman continuum flux $Q_0$ of an O7V star from Vacca \etal\ (1996) 
and assuming case B recombination for $T_e=$ 10000 K and $n_e =$ 100 cm$^{-3}$.},  
$N_{\rm O7V}$ plotted in the lower panel
of Fig.\ \ref{fig_plot_both}, ranges from $\sim$ 0.15 O7V stars (i.e.\ presumably 
corresponding to $\sim$ 1 late O or early B stars) to $\sim$ 400 O7V stars for
the brightest region.

\begin{figure}
\centerline{\psfig{file=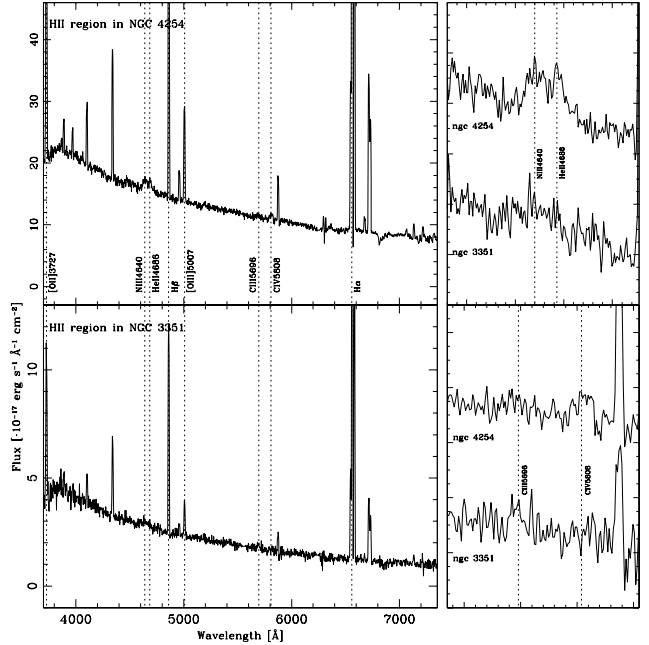,width=8.8cm}}
\caption{{\em Left panels:} FORS1/VLT spectra of a WR region in NGC 4254 (top)
and a candidate region in NGC 3351 (bottom) showing also the main line identifications.
{\em Right panels:} Zoom on the spectral region of the blue WR bump (top)
and the red WR bump (bottom) of the two spectra.}
\label{fig_spectra}
\end{figure}

Our search for WR features in metal-rich \hii\ regios proved quite
successful yielding with 27 WR detections a sample of unprecedented
size (cf.\ Castellanos 2001, Bresolin \& Kennicutt 2002).
The number of regions where different WR features were detected 
(hereafter called ``WR regions'')
at various levels of confidence are listed in Table \ref{tab_wr}.
The certain WR detections (defined as $\ge 2 \sigma$ detections) are listed in columns 2-4;
``candidate'' WR regions with emission line detections 1.1 $\le \sigma < $ 2 are given in cols.\
5-7. Visual inspection of the spectra yield essentially the same detection of the ``certain'' 
WR regions.
To illustrate the quality of our data sample spectra of a secure WR region and a 
candidate region are shown in Fig.\ \ref{fig_spectra}.

As also clear from Table \ref{tab_wr}, a large fraction of the WR regions
shows signatures of WR stars of both WN and WC types as anticipated from theoretical
expectations (Meynet 1995, SV98) and earlier studies of WR galaxies (Schaerer \etal\ 1997, 1999ab,
Guseva \etal\ 2000). 
In our sample $\sim$ 50 \% of WR regions show WC signatures;
predictions from the Meynet (1995) and SV98 models yield $\sim$ 30--77 \% at metallicities
$Z \sim$ 0.008 -- 0.040.
At least 1/3 of the WR regions harbour WC stars of late subtypes (WCL), characterised by their
strong \Ciii\ emission\footnote{It can be seen that \Civ\ is not formally detected
in all regions with \Ciii\ detections. However, in most of these cases (2 exceptions), 
\Civ\ is weak, but likely present.}.
The \Ciii/\Civ\ ratio indicates subtypes WC7 or WC8 assuming that the contribution of WN stars 
to \Civ\ is negligible; if this were not the case the mean spectral type could be of later
subtype.
So far relatively few WR ``galaxies'' (true starbursts or extra-galactic giant \hii\ regions)
with WCL stars are known (cf. Schaerer \etal\ 1999b).
However, as late WC types are expected to occur preferentially in metal-rich environments
(Smith \& Maeder 1991, Maeder 1991, Philipps \& Conti 1992) the high detection 
rate of \Ciii\ is not surprising.

The \ha\ luminosity distribution of the WR regions is shown in Fig.\ \ref{fig_plot_both}
(upper panel, dashed line). Clearly, WR stars are only detected in the brightest regions.
This is {\em not} due to the flux limit of our observations as can easily be seen
by comparison of the smallest WRbump fluxes 
($F({\rm WR})_{2 \sigma} \sim 4. 10^{-16}$ erg s$^{-1}$ cm$^{-2}$)
with the detection limit of the faintest emission lines 
($F_{\rm lim}(\hb) \sim 10^{-16}$ erg s$^{-1}$ cm$^{-2}$).
In fact our observations are essentially deep enough to allow in all galaxies 
\footnote{The few remaining ``candidate'' WR regions at high $L(\ha)$ are 
all located in NGC 4321, have formally a WR detection at $1.4 \la \sigma \la 2$,
and appear as rather clear WR detections at visual inspection. The observations
of NGC 4321 may be of less good quality due to limited seeing conditions.}
the detection of the blue WR bump of just $\sim$ 2--3 WNL stars,
assuming the average 4650-4686 \AA\ bump luminosity of a WN7 stars
of $10^{36.5}$  erg s$^{-1}$ (cf.\ Smith 1991, Schaerer \& Vacca 1998).
The number of WNL stars derived in this way is plotted in the lower panel of 
Fig.\ \ref{fig_plot_both} for regions with certain WR detections (filled squares)
and ``candidate'' WR regions (open circles).

As our detection limit allows for the detection of few ($\sim$ 2--3) average WNL stars, 
the subsample of the brightest regions with $F(\hb) \ga 5. 10^{-15}$ erg s$^{-1}$ cm$^{-2}$
could therefore represent a fairly complete sample of \hii\ regions 
``massive''/bright enough to allow a meaningful comparison between WR detections and
non-detections. 
However, a possible bias against regions with small \whb\ may exist (Sect.\ \ref{s_obs}).

In this subsample containing a total of 47 regions we find 20 objects without WR signatures,
or a fraction of $\sim$ 57 \% regions with WR signatures.
Such a high fraction of WR detections compares fairly well with the predictions
of 60--80 \% by Meynet (1995) and Schaerer \& Vacca (1998) using the high mass loss
stellar evolution tracks at metallicities $1/2.5 \la Z/\zsun \la 2$ 
for bursts with a standard Salpeter IMF and an upper mass cut-off \mup=120 \msun.
Given the fact that very young regions (ages 0 to $\sim$ 1.5--2 Myr) with
large expected \hb\ equivalent widths are notoriously absent (in the present sample and
other samples of \hii\ regions and galaxies) it is, however, not clear how significant
this finding is.

\section{Trends of WR populations with metallicity}
\label{s_wroh}

\begin{figure*}
\centerline{\psfig{file=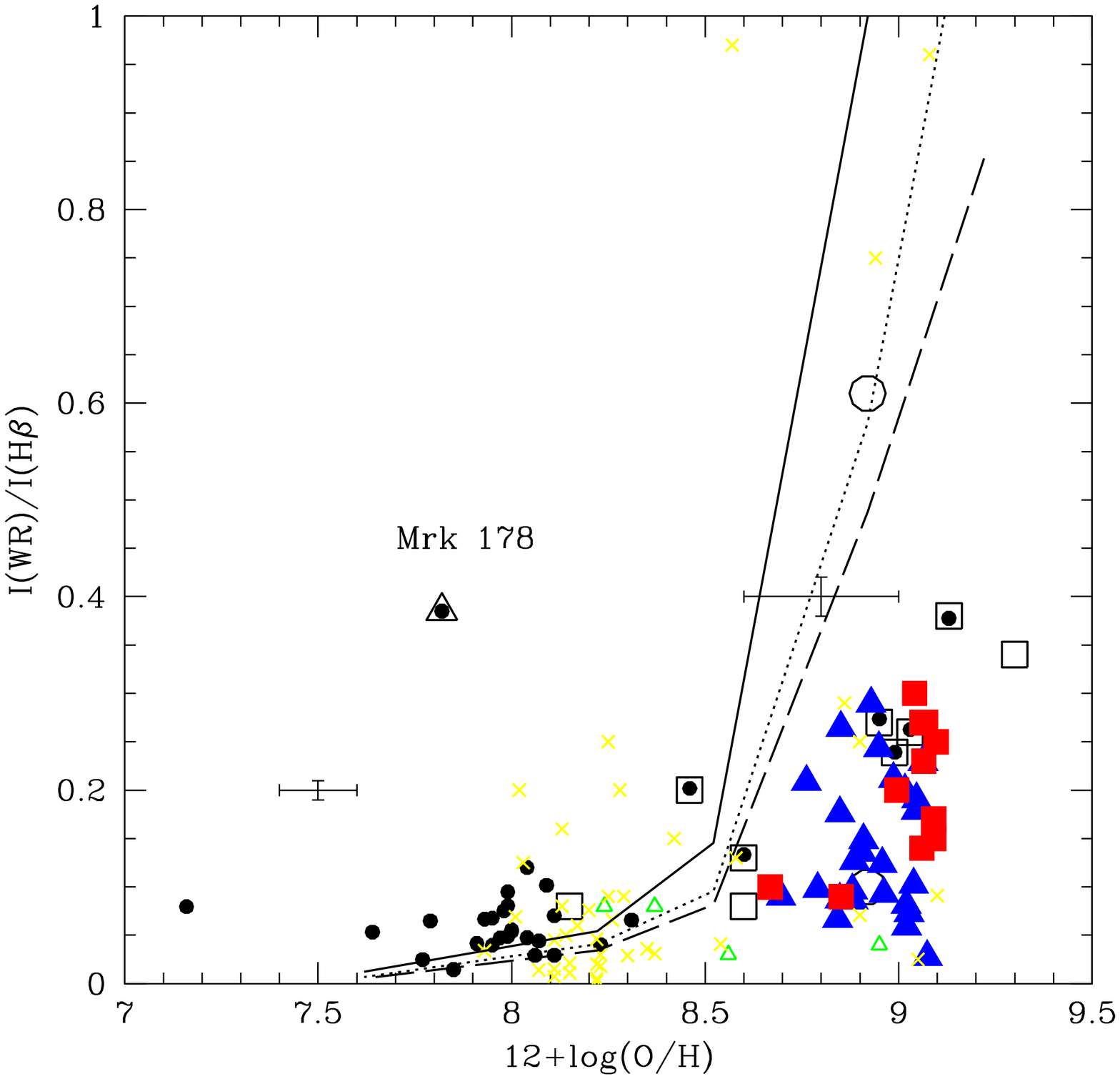,width=8.8cm}
            \psfig{file=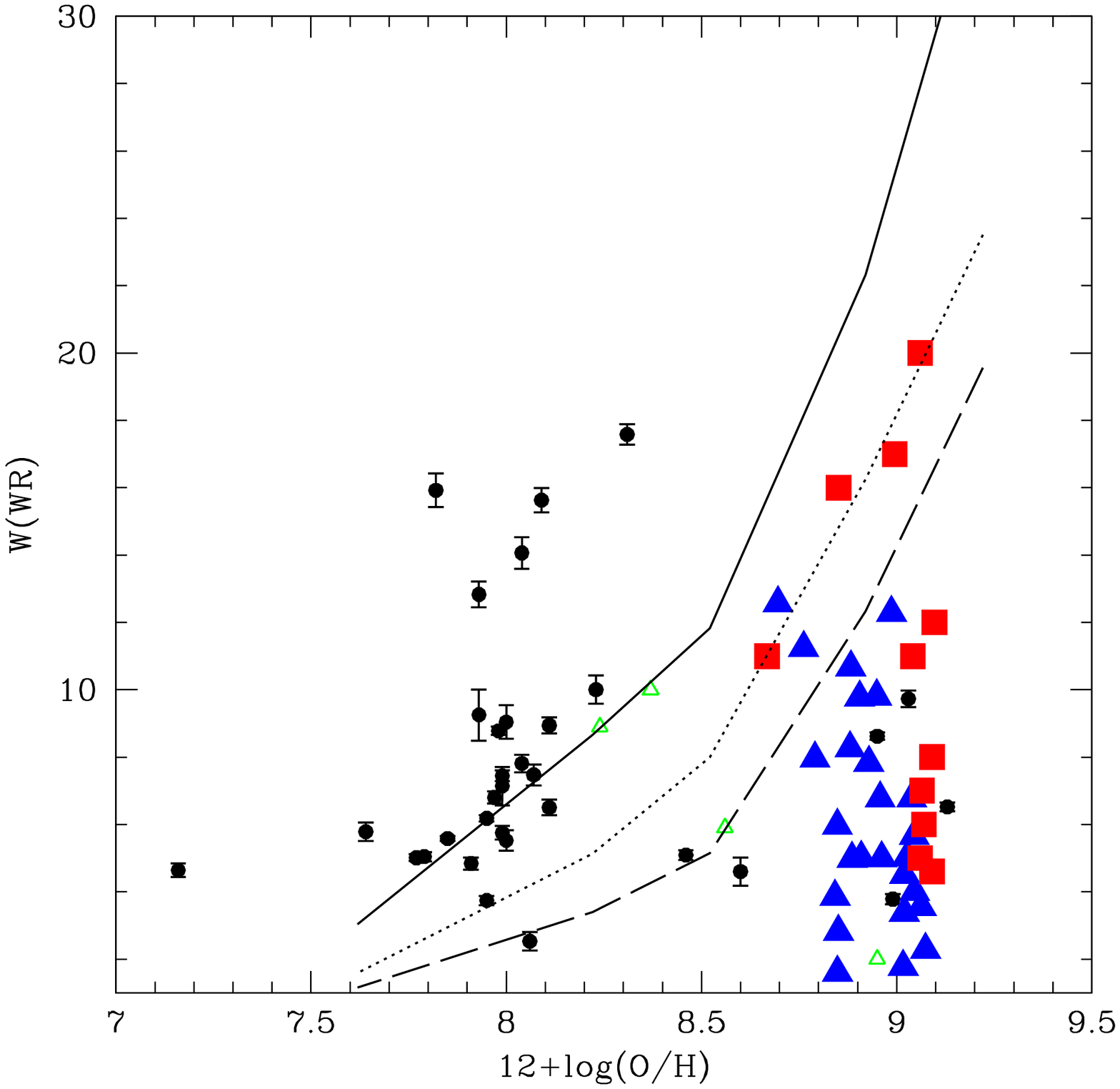,width=8.8cm}}
\caption{Observed WR-bump intensities (left panel) and equivalent widths (right panel)
as a function of
metallicity from the compilation of Schaerer (1999, crosses), the samples of Guseva 
\etal\ (2000, small filled circles), Castellanos \etal\ (2002a, smalle triangles),
SGIT00 (large open squares and circles),
BK02 (large filled squares) and the present data (large filled triangles).
Typical error bars for the Guseva \etal\ sample are shown.
{\em Maximum} predicted intensities and \wwr\ from the SV98 synthesis models are shown
for various star formation histories:
instantaneous bursts (solid), burst durations $\Delta t$ = 2 Myr (dotted)
and 4 Myr (long dashed). 
Note the overall agreement for \oh\ $\protect\la$ 8.5 and increasing discrepancy above.
See discussion in text.
}
\label{fig_bump_oh}
\end{figure*}

\subsection{Behaviour of the ``WR bump''}
Figure \ref{fig_bump_oh} shows the WR bump intensities and equivalent widths as a function
of metallicity for our metal-rich \hii\ regions (large filled triangles)
and the 11 WR regions in spiral galaxies recently reported by Bresolin \& 
Kennicutt (2002, large filled squares), together with data compiled by 
Schaerer (1999, small crosses) and Schaerer \etal\ (2000).

Our new measurements at high O/H are found to fill in the range from the previously observed 
maximum intensities/equivalent widths down to lower values.
Physically the maxima of \iwr\ and \wwr\ are expected to reflect the maximum WR/O 
star ratio achieved in bursts. 
No lower limit is expected; if present in a given sample, such a lower limit presumably 
reflects the detection limit of the WR features.

The increase of the upper envelope of \iwr\ with metallicity is known
since the work of Arnault \etal\ (1989) and has been reviewed by Schaerer (1999).
With few exceptions, max(\wwr) also seems to show an increase with O/H as shown 
here for the first time.
The increase of max(\iwr) is naturally interpreted as due to 
the increase of stellar wind mass loss with metallicity leading to lower minimum
mass limit for the formation of WR stars, $M_{\rm WR}$, thereby favouring the
presence of WR stars at high metallicity (cf.\ Maeder \etal\ 1981, Arnault \etal\ 1989,
Maeder 1991). Other effects, e.g.\ a lowering of the \hb\ flux due to
a) increasing amounts of dust absorbing ionising radiation or b) lower average stellar
temperatures at high O/H due to modified stellar evolution, could also
play a role (cf.\ Schaerer 1999), but are likely secondary.

The maxima of the predicted WRbump intensities and equivalent widths
computed with the code of SV98 
with a ``standard'' Salpeter IMF for instantaneous bursts (solid line),
and extended bursts of duration  $\Delta t$ = 2 Myr (dotted),
and 4 Myr (long dashed) are overplotted on Fig.\ \ref{fig_bump_oh}.
As already shown earlier (cf.\ Schaerer 1996, 1999, Mas-Hesse \& Kunth 1999,
Guseva \etal\ 2000) the range of observations at subsolar metallicities
(\oh\ $\la$ 8.6) is fairly well reproduced by the models,
when accounting for the various uncertainties (e.g.\ missing \hb\ flux
in slit observations, some objects with small numbers of WR stars, some
poor spectra; cf.\ discussion in Guseva \etal).
The new sample of metal-rich objects plotted here shows
WRbump strengths smaller than the maxima predicted by the ``standard'' models.
The possible reasons for this behaviour are discussed in 
Sect.\ \ref{s_models} where detailed model comparisons are undertaken.

\begin{figure}
\centerline{\psfig{file=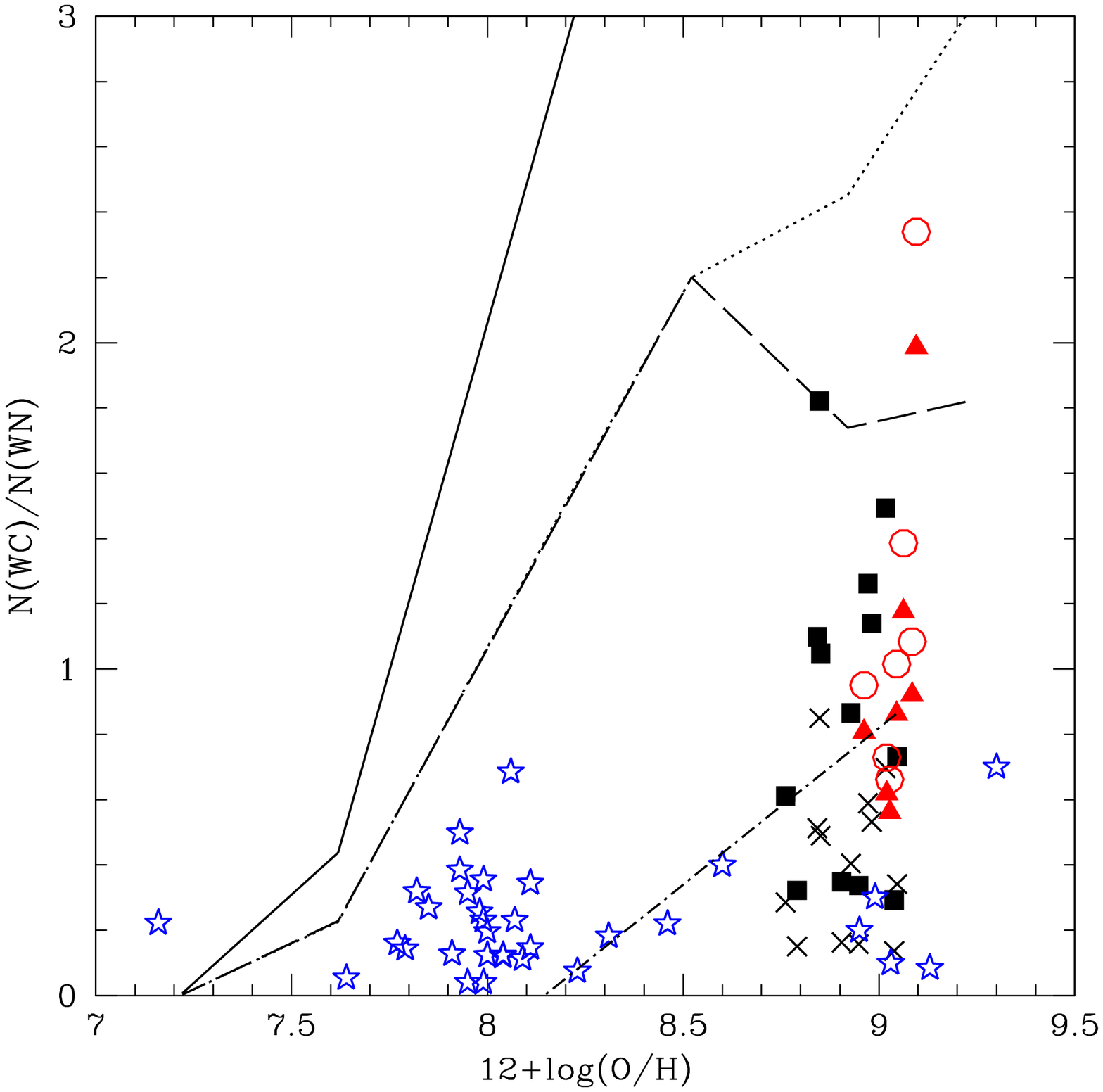,width=8.8cm}}
\caption{Estimated number ratio of WC/WNL stars versus metallicity.
Data derived from \Civ/WRbump observations from the sample of Guseva \etal\ 
(2000) and SGIT00 are shown by stars; the assumed mean spectral type of the WC stars
is WC4 for \oh\ $<$ 8.4 and WC7 for higher metallicities.
Results from our VLT data are shown for different assumptions:
(1) from \Civ/WRbump assuming a WC7 spectral type (filled squares), or WC4 (crosses).
(2) from \Ciii/WRbump assuming WC7 spectral type (filled triangles), or WC8 (open cirles).
Maximum predicted WC/WN ratios from the SV98 models are shown for 
instantaneous bursts (solid), and burst durations $\Delta t$ = 2 Myr (dotted),
and 4 Myr (long dashed). 
The observed trend of WC/WN with metallicity in Local Group galaxies,
thought to represent an average for constantly star forming regions,
from Massey \& Johnson (1998) is shown by the dash-dotted line.
}
\label{fig_wcwn_oh}
\end{figure}

\subsection{WC/WN ratio}
We have estimated the relative number ratio of WC and WN stars, shown in Fig.\
\ref{fig_wcwn_oh}, in several ways.
First the number of WN stars, $N({\rm WNL})$ assuming late WN subtypes dominate, 
is derived from the luminosity of the blue WR bump, as described above. 
The number of WC stars, $N({\rm WC})$, is estimated from the \Civ\ or \Ciii\ luminosity
where measured, again assuming that WN stars do not contribute to these lines.
As the observed average luminosity of WC stars in these lines varies strongly
with subtype (see SV98), the estimated $N({\rm WC})$ depends on the assumption of the
dominant WC subtype. As the observations (see above, Guseva \etal\ 2000, Schaerer
\etal\ 1999a) indicate that early types ($\sim$ WC4) dominate at low metallicity, while
WC7-8 dominate at high \oh, we assume these mean WC subtypes for the sample
of Guseva \etal\ (2000).
For our high metallicity sample, the estimated $N({\rm WC})/N({\rm WNL})$ ratios is estimated 
adopting different assumptions on the WC subtype and using \Civ\ or \Ciii\
(see Fig.\ \ref{fig_wcwn_oh}).

The resulting estimates show a fairly clear trend of an increasing upper envelope 
for $N({\rm WC})/N({\rm WNL})$ with metallicity. 
Furthermore, and in contrast with the limited sample of Guseva \etal\ (2000),
we now find at the high metallicity end a number of objects with 
$N({\rm WC})/N({\rm WNL}) \ga$ 0.5--1. and a WC/WN number ratio larger than the observed trend 
in Local Group galaxies by Massey \& Johnson (1998), indicated by the dash-dotted
line in Fig.\ \ref{fig_wcwn_oh}.
Indeed, while the regions observed by these authors are thought to correspond to
averages large enough to represent the equilibrium $N({\rm WC})/N({\rm WNL})$ value at constant
star formation, larger (and obviously also smaller) values should be found in 
regions with fairly short bursts.

A more quantitative interpretation of the observed WC to WN ratio appears difficult
for the following reasons. First the uncertainties in the estimated $N({\rm WC})/N({\rm WNL})$ are
quite large (cf.\ above); second, detailed evolutionary synthesis model predictions 
of $N({\rm WC})/N({\rm WNL})$ depend quite strongly on the adopted interpolation techniques
(cf.\ SV98, comparison between results from SV98 models and {\em Starburst99}
(Leitherer \etal\ 1999), also Massey 2002); third, other comparisons with synthesis
models reveal potential difficulties (cf.\ below).
In any case the SV98 models predict the maximum WC/WN number ratios
indicated in Fig.\ \ref{fig_wcwn_oh} by the solid line for instantaneous bursts,
and burst durations of $\Delta t=$ 2 Myr (dotted) and 4 Myr (dashed) respectively.

\section{Detailed comparison of WR populations with synthesis models}
\label{s_models}

\subsection{Procedure}
\label{s_procedure}
To interpret quantitatively the observational data we use evolutionary synthesis models and
proceed essentially as in SGIT00.
The following main observational constraints are used:
\begin{enumerate}
\item {\em \hb\ and \ha\ equivalent widths.} The former is used as a primary 
  age indicator; once \whb\ is reproduced $W(\ha)$ may serve as an independent
  consistency test for the predicted spectral energy distribution (SED) in the red
  (cf.\ SGIT00).
\item {\em Nebular line intensities.} $F$(\ha)/$F$(\hb) determines the extinction
  of the gas. The use of other line intensities requires detailed photoionization
  modeling which is beyond the scope of this paper.
\item {\em Intensities and equivalent widths of the main WR features.}
  The blue bump and \Civ\ (red bump)
  serve as main constraints on the WR population. To avoid uncertainties
  in deblending individual contributions of the blue bump we prefer to
  use measurements for the entire bump. 
  In contrast to the spectra of metallicity objects our spectra show 
  no evident contamination from nebular lines (e.g.\ [Fe~{\sc iii}] $\lambda$ 4658,
  nebular \heii).

  To potentially disentangle between various effects (underlying ``non-ionizing'' 
  population, loss of photons, differential extinction between gas and stars)
  it is important to use both equivalent widths and relative \iwr\ intensities
  (cf.\ Schaerer \etal\ 1999a).
\end{enumerate}

For the model comparisons we use calculations based on the evolutionary synthesis code 
of SV98, which in particular includes the most recent calibration of WR line
luminosities used to synthesize the WR features, up-to-date stellar tracks, 
{\em CoStar} stellar atmospheres for O stars, pure H-He models for WR stars 
and Kurucz models for cooler stars
(see SV98 for a full description). 
Except for the improved O star atmospheres used by SV98 the {\em Starburst99} 
synthesis models (Leitherer \etal\ 1999) use the same basic input physics. 
New generation
stellar atmosphere models for O and WR stars including a full treatment of non-LTE 
line blanketing and stellar winds 
have just now become available for the 
use in synthesis models (Smith \etal\ 2002).
However, as the quantities of interest here depend only on the total number of
Lyman continuum photons which is not altered, the use of these more sophisticated 
atmosphere models does not affect our results.

It is important to stress that in all cases the high-mass loss stellar tracks
of Meynet \etal\ (1994) are used.
It is thought that this adjustment of mass-loss, treated  like a free parameter,
will become ultimately obsolete when a proper treatment of the various effects
of stellar rotation is made in the stellar evolution models.
First results tend to indicate that this may indeed be the case
(Meynet 1999).
The Meynet \etal\ (1994) tracks are chosen as they reproduce a large number of 
properties of individual WR stars and WR populations (including especially 
relative WR/O ratios for  a standard Salpeter IMF) in Local Group galaxies 
(Maeder \& Meynet 1994).
The use of other tracks (e.g.\ the ``normal'' mass loss tracks)
which are known to disagree with these basic constraints on WR and O star populations,
would imply a strong inconsistency with the Local Group data.

\begin{figure*}
\centerline{\psfig{file=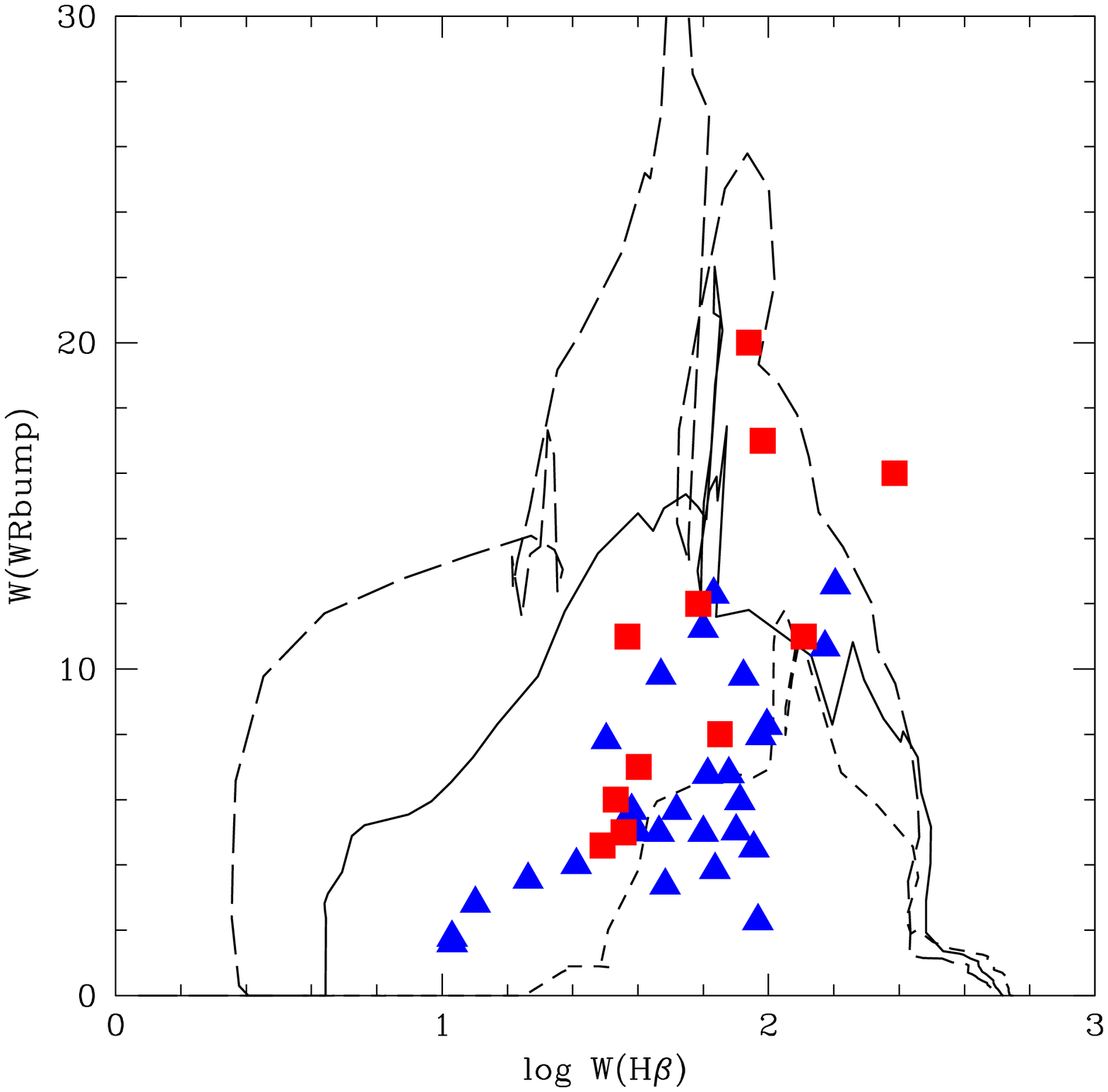,width=8.8cm}
            \psfig{file=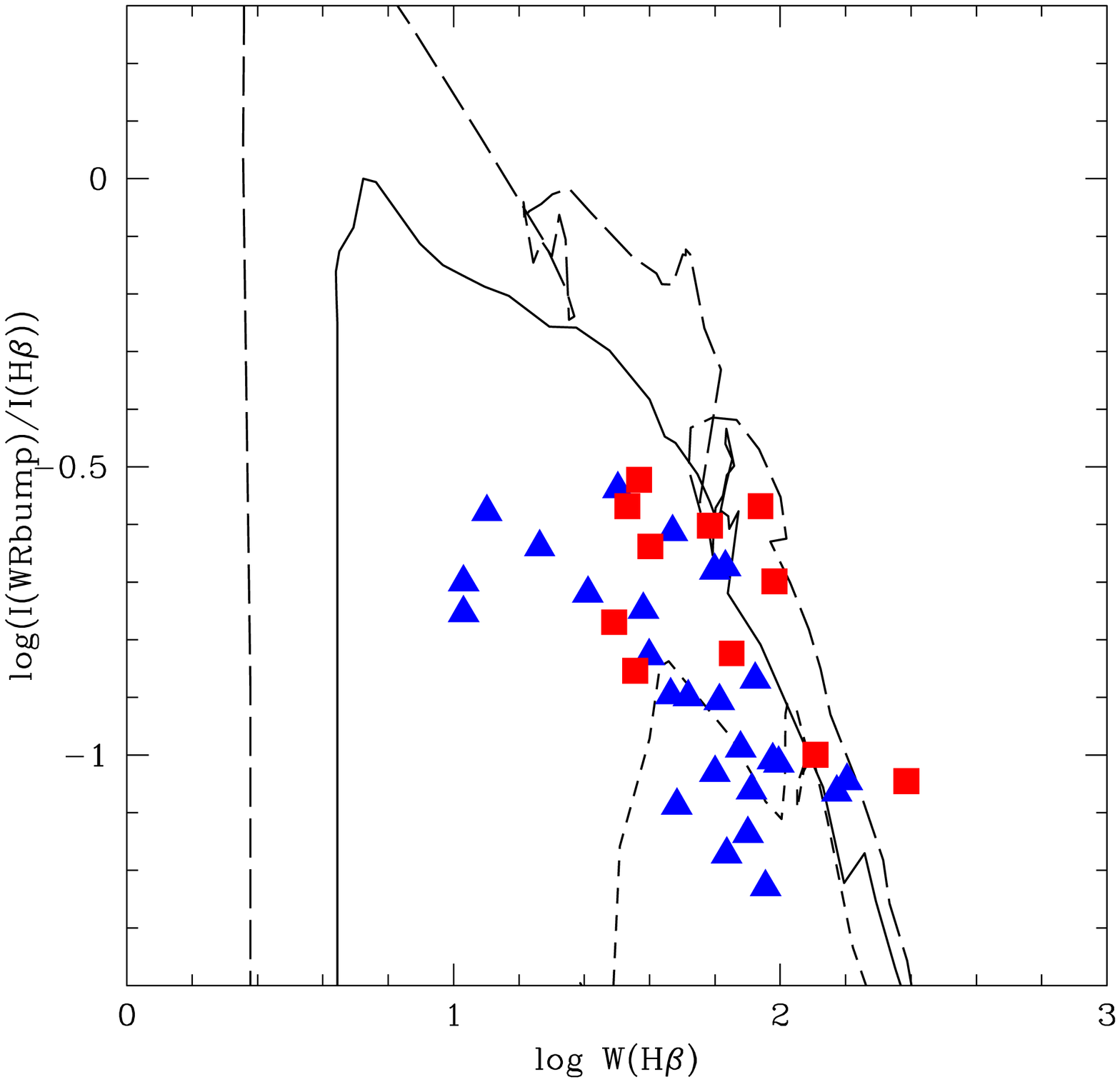,width=8.8cm} }
\caption{Observed and predicted equivalent width (left panel) and line intensity
with respect to \hb\ (right panel) as a function of \whb. 
Our VLT sample is shown by (black) triangles, the BK02 sample with (red) squares.
Typical uncertainties are 5--10 \% for \whb, $\le$ 10 \% for $W($WRbump$)$,
and $\sim$ 0.05 dex in $\log(I($WR)/$I($\hb)).
Model predictions are shown for instantaneous bursts with ``standard'' IMFs
at  $Z$=0.008 (dashed line), $Z$=\zsun=0.02 (solid line), and $Z$=0.04 (long dashed line).
Note the overprediction of the WRbump strength in high metallicity models compared
with the observations.
}
\label{fig_std}
\end{figure*}

The basic model parameters we consider are:
\begin{enumerate}
\item[a)] {\em Metallicity.} Stellar tracks covering metallicities 
  $Z=$, 0.008, 0.02 (solar), and 0.04.
\item[b)] {\em IMF slope and upper mass cut-off (\mup).} We adopt a Salpeter IMF
  (slope $\alpha$=2.35), and \mup=120 \msun\ as our standard model.  
\item[c)] {\em Star formation history (SFH).} Models for instantaneous bursts
  (coeval population), extended burst durations (constant SF during period \dt; 
  in this case age=0 is defined at the onset of SF, i.e.\ corresponds to that of
  the oldest stars present),
  and constant SF are considered. 
\item[d)] {\em Fraction of ionizing Lyman continuum photons ($f_\gamma$).} 
   $f_\gamma$ indicates the fraction of ionizing photons absorbed by the gas.
   Our standard value is $f_\gamma=1$.
   Values $f_\gamma < 1$ are used to simulate various effects (e.g.\ dust absorption, photon
   leakage outside regions, etc.) leading to a reduction of photons available for
   photoionization.
\end{enumerate}

Unless stated otherwise our models are calculated assuming an IMF fully 
sampled over the entire mass range (as in SV98).
For some cases we have also done model calculations
based on a Monte Carlo sampling of the IMF, in order to quantify the effects
of statistical fluctuations due to the finite number of massive stars.
We have verified our calculations by comparison with the Monte Carlo models and  
analytical results of Cervi\~no \etal\ (2000, 2002).

\begin{figure*}
\centerline{\psfig{file=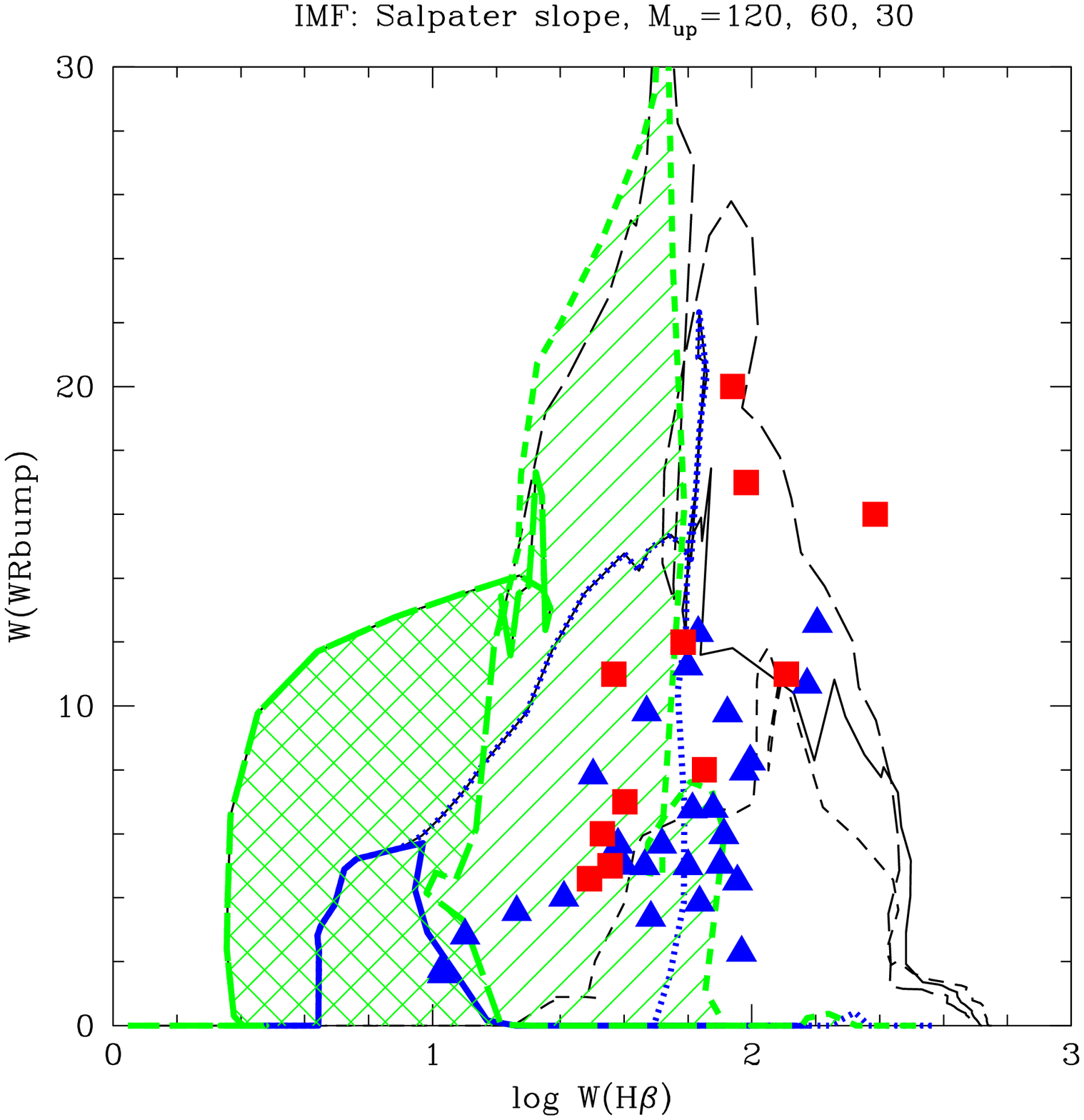,width=8.8cm}
            \psfig{file=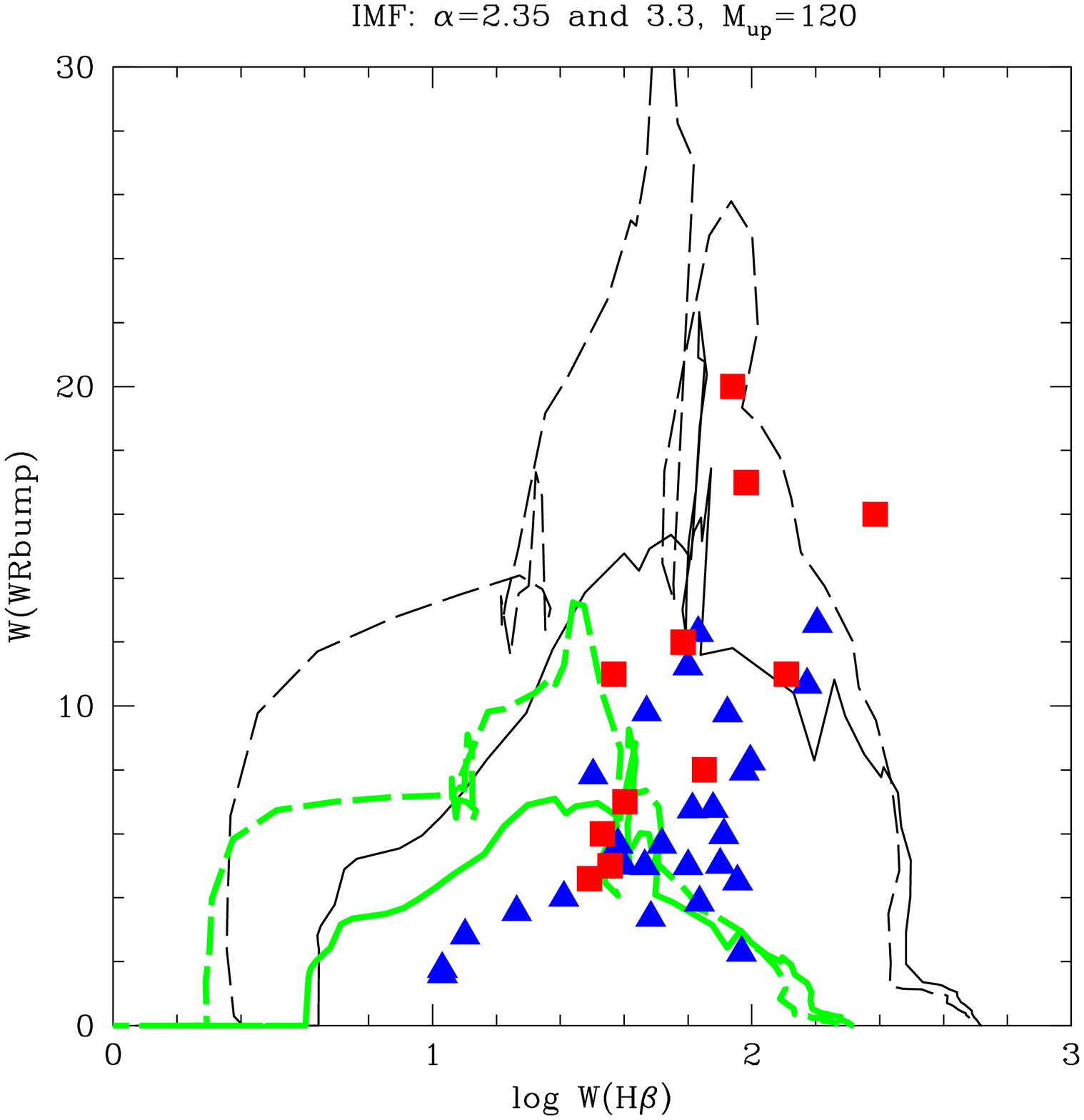,width=8.8cm}
            }
\caption{Observed and predicted WRbump equivalent width as a function of \whb.
Standard model predictions are shown for instantaneous bursts at
$Z$=\zsun=0.02 (solid line), and $Z$=0.04 (long dashed line).
{\em Left panel:}
Thick (green) lines with the same styles show models with a standard IMF slope
($\alpha=2.35$) and upper mass cut-offs of \mup=60 and 30 \msun\
delimiting the singly and doubly shaded regions respectively.
{\em Right panel:} 
Thick (green) lines with the same styles show models with a IMF slope
of $\alpha=3.3$ and \mup=120.}
\label{fig_alpha}
\end{figure*}

\subsection{Results}
\label{s_result}

A comparison of the observed equivalent widths and relative intensity of the
WR bump with standard model predictions at different metallicities is presented
in Fig.\ \ref{fig_std}. The following points can be seen from this
Figure:
\begin{itemize}
\item The observed trends of decreasing (increasing) equivalent width (line intensities)
  of the WR features with decreasing \hb\ equivalent width agree 
  with the instantaneous burst model predictions over the same \whb\ range.
  Though shorter, part of the initial phase of increasing (decreasing) \wwr\ (\iwr)
  corresponding to the onset of the WR rich phase does not seem to be detected.

\item Most importantly, {\em essentially all the observed WR features are weaker 
   than the predictions of our standard models for instantaneous bursts at solar 
   or higher metallicity.}
   Very similar observational trends are also found in the metal-rich \hii\ region
   samle of BK02 and the metal-rich starbursts of Schaerer \etal\ (2000).
   This is in stark contrast with observations at lower metallicity 
   where generally an good agreement is found with short burst models (e.g.\ 
   Schaerer \etal\ 1999a, Guseva \etal\ 2000).

\item The above result is found independently from the observed WR equivalent widths
  and relative line intensities WRbump/\hb. This indicates that the discrepancy between
  models and observations is not related to the possible presence of an underlying
  older stellar population, which would dilute (reduce) equivalent widths but
  not alter the relative line intensities.
\end{itemize}

\begin{figure}
\centerline{\psfig{file=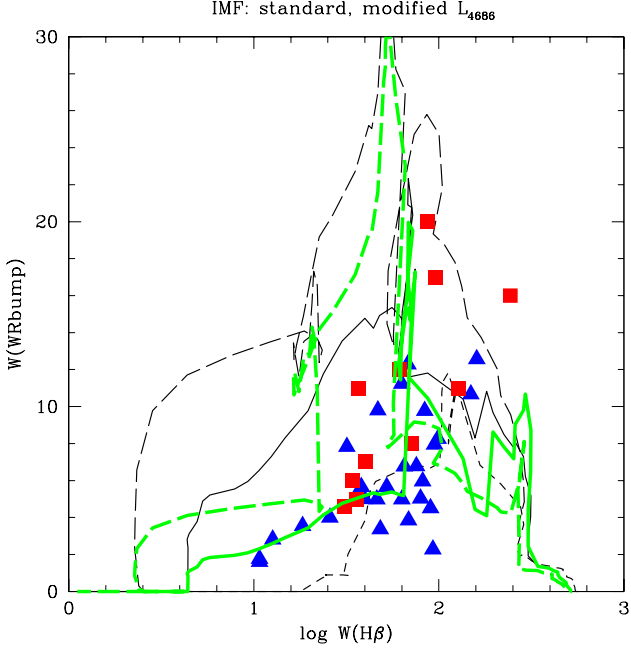,width=8.8cm}}
\caption{
Observed and predicted WRbump equivalent width as a function of \whb.
Standard model predictions are shown for instantaneous bursts at
$Z$=\zsun=0.02 (solid line), and $Z$=0.04 (long dashed line).
Thick (green) lines with the same styles show the predictions with the
modified  $L_{\rm 4686}$ calibration for WNL stars leading to an important
reduction of the WRbump for $\log \whb  \protect\la 1.8$, due to the lower average
luminosity of WNL stars in bursts with ages $ \protect\ga$ 4--5 Myr (for $Z=$ 0.02).
}
\label{fig_wnl_mod}
\end{figure}

The following possibilities (one or a combination thereof) could be invoked to 
explain the discrepancy between our observations and models:
\begin{enumerate}
\item {\em The metallicities of our HII regions are overestimated.} 
  Indeed the observations could be reconciled with burst models with a 
  ``standard'' IMF for metallicities $Z \sim$ (1/2.5--1) \zsun, as shown in the left
  panel of Fig.\ \ref{fig_std} (short dashed line).
  However, despite the uncertainties in the O/H determinations (cf.\ Sect.\ 
  \ref{s_props}) such low average metallicities seem very implausible.

\item {\em Extended bursts.} Such a scenario has been invoked by SGIT00 for the 
sample of metal-rich starbursts based on the finding of red supergiant features
in their spectra and the fact that these distant objects are mostly nuclear starbursts
observed through apertures corresponding to relatively large spatial scales.
In this case all observed properties could quite well be fitted with ``standard'' 
solar metallicity models for burst durations $\Delta t \sim$ 4--10 Myr.
However, in view of the different nature (disk \hii\ regions) of the present sample,
indications of relatively short formation time scales of \hii\ regions 
(e.g.\ Massey \etal\ 1995), 
and the lack of direct signatures of older/red supergiant populations (cf.\ below)
it seems quite unjustified to appeal to extended burst to solve the observed discrepancy.

\item {\em A modified IMF (upper mass cut-off and/or slope).}
In a plot like Fig.\ \ref{fig_std}, a Salpeter IMF with a lower upper mass 
cut-off simply implies that the curve plotted here (for \mup=120\msun) is joined
at lower \whb\ as the WR stars from the most massive stars are absent.
This is illustrated for the cases of \mup\ $=$ 30 and 60 \msun\ by the shaded domains
in Fig.\ \ref{fig_alpha}.
The shape of the predicted WR equivalent width or line intensity remains, however,
unchanged. Therefore the observed discrepancy cannot be resolved with an
IMF of Salpeter slope and a lower value of \mup\ (see also Sect.\ \ref{s_imf}).

Models with steeper, variable IMF slopes ($2.35 < \alpha \la 3.3$) and \mup\
$\sim$ 60--120 \msun\ could reproduce most of the objects, with the exception 
of the lowest \whb\ objects (see Fig.\ \ref{fig_alpha}).
As the least metal-poor objects in our sample are probably of similar nature
as young clusters or \hii\ regions in our Galaxy whose stellar content has 
been studied in detail, we may presume that their IMF (slope and \mup) should 
be similar.
Since none of the Galactic regions have shown convincing evidence of a 
strong deviation of the IMF slope from the Salpeter value (see Massey 1998 
and references therein),
we think that such a steeper slope is an unlikely explanation.

\item {\em Incorrect stellar evolution models and/or ``calibration data'':}
Although the adopted tracks (Meynet \etal\ 1994) compare fairly well with 
various observations, several failures of the non-rotating stellar models are also
known (see e.g.\ Maeder 1999).
However, the used tracks have essentially been calibrated/adjusted to fit the 
observed WR/O ratio in various regions of our Galaxy and Local Group objects 
which are though to be at equilibrium, i.e.\ showing relative populations 
corresponding to constant star formation (see compilation in Maeder \& Meynet 1994).
The relative WR/O star ratio is the one most directly related to our (time resolved) 
observables. 
As this calibration yields a fairly good agreement over a large metallicity 
range ($1/10 \la Z/\zsun\ \la 2$) 
there seems little room for changes in the tracks which could reduce the predicted 
WRbump by the required factor of $\sim$ 2 without violating the WR/O constraints 
in the Local Group.

One could argue that the calibration data, the observed WR/O number ratio
at solar metallicity and above could be incorrect due to possible
incompleteness or biases in the stellar counts (see e.g.\ related discussions
in Massey \& Johnson 1998).
However, to reconcile our WR observations in \hii\ regions with the corresponding
counts for our Galaxy and M31 would require a downward revision of the
relative WR/O ratio by up to a factor of 2, which seems highly unlikely.

\item {\em Uncertainties in synthesis of the WR bump:}
Presently the calculation of observables related to WR stars is simply done in 
the following way in evolutionary synthesis models.
The different emission line strengths are computed by multiplying the predicted 
number of WR stars (grouped in different types and/or subtypes) with their
average line luminosity as derived from observations of a sample of WR stars
(see SV98).
Interestingly the intrinsic line luminosity of the strongest line of the WR bump, 
\Heii, shows a rather large scatter, namely 
$L_{\rm 4686}=(1.6 \pm 1.5) \times 10^{36}$ erg s$^{-1}$ in the Galactic and LMC 
WNL calibration sample of SV98 with a possible increase of $L_{\rm 4686}$ with 
the stellar bolometric luminosity $L$ (see Fig.\ 1 of SV98).
Such a luminosity dependence of $L_{\rm 4686}$ with $L$ could in fact 
(partly or fully) explain the observed discrepancy as we will now show. 

Splitting the WNL calibration in two domains with luminosities above/below 
$\log L/\lsun = 6$, SV98 found average line luminosities 
$L_{\rm 4686} = 5.6 \times 10^{35}$ ($\log L < 6$) and  
$L_{\rm 4686} = 3.1 \times 10^{36}$ ($\log L > 6$).
Replacing in the synthesis models the overall average for WNL stars by these 
quantities leads to an important reduction of \wwr\ in solar metallicity bursts with
\whb\ $\la$ 60--70 \AA, as shown
in Fig.\ \ref{fig_wnl_mod}\footnote{At $Z=$0.04 one has a reduction
of \wwr\ both in the early burst phase (ages $\la$ 3.5 Myr) at $\log \whb\ \ga$ 1.8 \AA\
and after $\ga$ 6 Myr ($\log \whb\ \la$ 1.35), while in between the models
predict that WC stars dominate the bump (cf.\ SV98, Figs.\ 2 and 11).}.
At larger \hb\ equivalent widths (corresponding to ages $\la$ 4--5 Myr for
$Z=$0.02) the WRbump predictions are less modified, since {\em a)} 
WC stars contribute more importantly to the bump and {\em b)} only the youngest
bursts with very high \whb\ are dominated by very luminous WNL stars showing
thus larger \wwr.

\end{enumerate}

The last option (5) seems the most likely explanation to explain the 
surprisingly low WR equivalent widths and intensities in our sample of 
metal-rich \hii\ regions.
Implications on earlier studies of
WR galaxies are briefly discussed in Sect.\ \ref{s_discuss}.

\begin{figure}
\centerline{\psfig{file=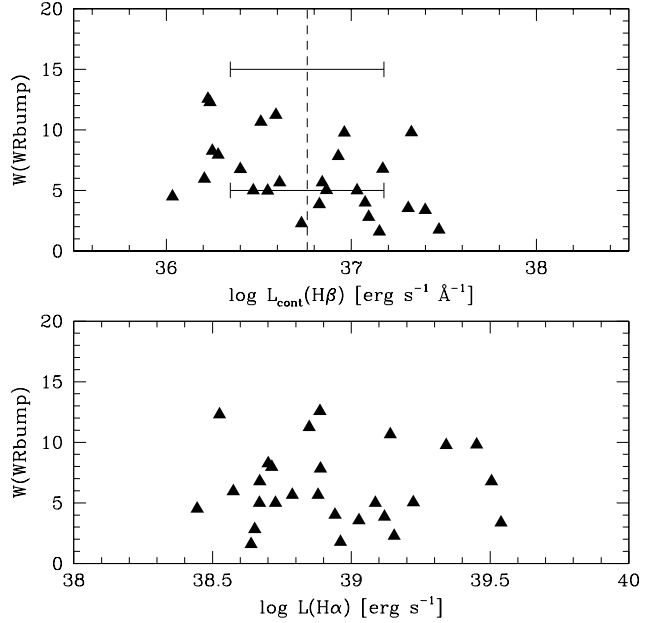,width=8.8cm}}
\caption{Equivalent widths of the WRbump as a function of the monochromatic continuum
luminosity at \hb\ in erg s$^{-1}$ \AA$^{-1}$ (top panel), and as a function of the 
\ha\ luminosity for the WR regions of our sample.
The mean and dispersion (1 $\sigma$) of $\log L_{\rm cont}(\hb)$ is plotted in the 
top panel.}
\label{fig_abs_scales}
\end{figure}

\begin{figure}[ht]
\centerline{\psfig{file=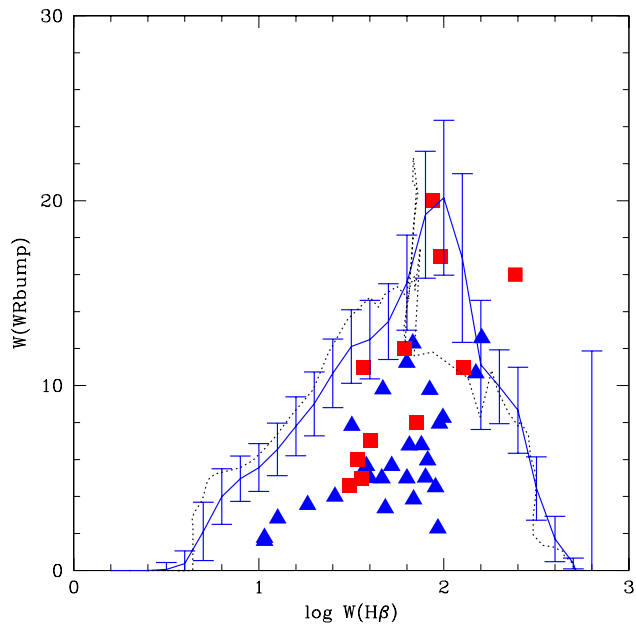,width=8.8cm}}
\caption{Same as Figs.\ \ref{fig_std} (left) and \ref{fig_wnl_mod}
showing the comparison between model predictions using a fully sampled, analytical
IMF (dotted line, black) and the predicted mean \wwr\ and $1 \sigma$ scatter
(solid line, blue) for a burst of scale/mass corresponding to the average observed
continuum luminosity. Observations are shown using the same symbols and in the previous
Figs. Note, the deviation of the mean values for $\log \whb \sim$ 1.9--2.2
is due to a numerical artifact.
The comparison shows that no significant bias is expected and that the
scatter is too small to resolve the discrepancy with observations.}
\label{fig_mc_wrbump}
\end{figure}

In contrast, the following hypothesis or effects altering observed equivalent widths 
and/or relative line intensities {\bf cannot} be the cause of the discrepancy:
\begin{itemize}
\item {\em Underlying (old) populations diluting the \wwr\ measurements.}
  The line intensities are unaffected by underlying populations and show the same
  discrepancy as \wwr\ (cf.\ above). 
  Furthermore, inspection of our spectra show little or no indication of
  signatures of older stellar populations.
\item {\em Differential extinction between gas and stars} as frequently observed
  in \hii\ galaxies and starbursts (cf.\ Fanelli \etal\ 1988, Calzetti 1997,
  Mas-Hesse \& Kunth 1999, SGIT00). If present such an effect alters \whb\ and
  \iwr\ but not \wwr. To bring the observations to agreement with standard models 
  would require that the gas suffered a {\em lower} extinction than the stars
  (implying thus lower corrected \whb\ and larger \iwr), opposite to what is
  found empirically.
\item {\em Leakage of ionising photons outside the observed regions, dust absorption
  of ionising photons}, or other mechanisms reducing the fraction $f_\gamma$
  of Lyman continuum photons used for photoionisation. Correcting the observations
  for such an effect would increase the observed \whb\ and decrease \iwr,
  which does not reconcile observations and theory (see Fig.\ \ref{fig_std}). 
\item {\em Effects due to varying seeing conditions and limited slit widths} could  
lead to a loss of nebular emission in the aperture or a fraction of the stellar 
light. The former was just discussed (``leakage''). If the WR distribution were 
systematically more extended than the other stars responsible of the continuum, 
this effect could lead to reduced WR bump equivalent widths.
No such trend between \wwr\ and the seeing conditions is found.

\item {\em Stochastical fluctuations of the IMF} due to small number statistics 
  of the massive stars, as invoked by Bresolin \& Kennicutt (2002).
  Although indeed expected to introduce some scatter (see Cervi\~no \etal\ 2000, 2002)
  this effect cannot explain the discrepancy for several reasons. 
  First, the sample discussed here is sufficiently large and shows clear 
  observational trends with relatively small scatter. 
  In addition, no systematic trend of \wwr\ (and \iwr) with absolute
  scale, such as measured by the continuum luminosity or \ha\ luminosity,
  is observed in our sample as shown in Fig.\ \ref{fig_abs_scales}.
  Second, Monte Carlo simulations we have undertaken for cluster sizes 
  corresponding roughly to 
  the observed average continuum luminosity of $<\log L_{\rm cont}> \sim 36.8 \pm 0.4$
  \footnote{The model $L_{\rm cont}$ at young ages ($\la$ 1 Myr) is scaled to the observed
  value, implying a stellar mass of $\sim 5. (3.3) \times 10^4$ \msun\ for
  a Salpeter IMF between 0.8 (2.) and 120 \msun. Accounting for somewhat older ages
  would yield a slightly larger mass.} predict a typical relative scatter 
  of $\sigma(W({\rm WRbump}))/\wwr \sim$ 25 \%\ (cf.\ Cervi\~no \etal\ 2002) 
  -- too small to account for the discrepancy -- and no significant bias towards
  lower values as illustrated in Fig.\ \ref{fig_mc_wrbump}. 
   
\item The use of different stellar tracks, such as e.g.\ the Geneva tracks with
standard mass loss tracks adopted by Bresolin \& Kennicutt (2002) for the 
Monte Carlo models, which {\em do not} reproduce basic constraints of massive stars
and stellar populations in Local Group objects 
cannot be a solution as they would lead to important inconsistencies (cf.\ 
discussion in Sect.\ \ref{s_procedure}).

\end{itemize}

\subsection{Discussion}
\label{s_discuss}

In Section \ref{s_result} we have argued that, compared to the normal 
prescription used in our SV98 synthesis models, a different prescription
should preferrably be adopted to predict more accurately the \Heii\ 
emission from WN stars.
As several earlier studies including ours (e.g. Schaerer 1996, 1999,
Schaerer \etal\ 1999a, Guseva \etal\ 2000, SGIT00) are based on the 
use of the simple average \Heii\ line luminosity of SV98 for WNL stars,
it is important to assess if or to what extent the use of a luminosity dependent
prescription would affect the results from previous studies.

To verify this we have recomputed several sets of models for sub-solar metallicities.
The maxima of the WRbump intensity and \wwr\ (cf.\ Fig.\ \ref{fig_bump_oh})
are only slightly modified (increased at \oh\ $\la$ 8.5, and decreased above) 
and lead to a somewhat smaller increase with O/H, improving the agreement
with the observations.
For metallicities $Z \la$ 1/2 \zsun\ the predicted WR bump is found to be 
larger at all ages (as the bulk of WN stars are of high luminosity), 
whereas for higher metallicities both larger/smaller WRbump strengths
are predicted depending on the burst age (\whb), as for the cases shown
in Fig.\ \ref{fig_wnl_mod}.
These changes improve the comparison with observations at low $Z$ (see e.g.\
Fig.\ 7 of Guseva \etal\ 2000). No clear statement can be made for intermediate 
metallicities. 
A better understanding of the dependence of the WR emission lines on the stellar 
parameters appears necessary to improve the accuracy of the predictions
of WR features in evolutionary synthesis models.
The impact of newly available stellar evolution models including the effects
of rotation on interior mixing and mass loss on massive star populations
remains also to be explored.

\section{Constraining the upper end of the IMF}
\label{s_imf}
As shown above, and in contrast to previous studies considering WR populations 
in objects with sub-solar metallicities, the quantitative modeling of 
the WR features in metal-rich \hii\ regions is not straightforward.
The reliability of constraints on the IMF obtained from modeling of the
WR features appears thus unclear at present and other constraints are 
therefore desirable.

\begin{figure}[ht]
\centerline{\psfig{file=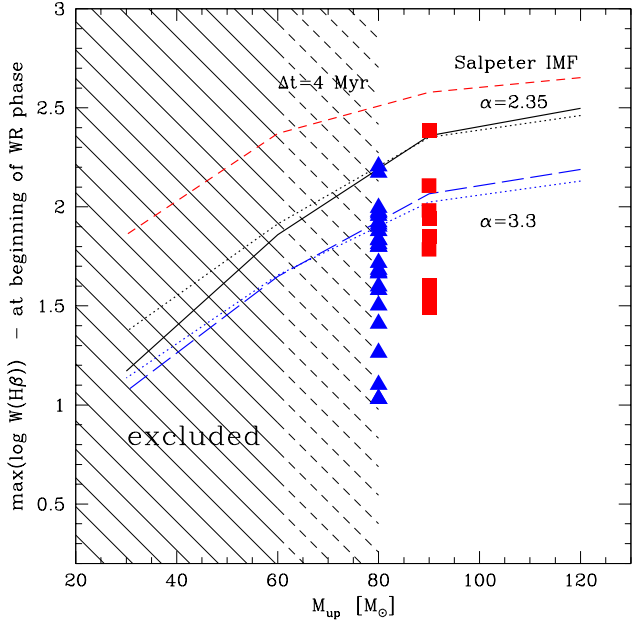,width=8.8cm}}
\caption{Maximum predicted \hb\ equivalent width at the beginning of 
the WR phase as a function of \mup\ for solar metallicity ($Z=0.02$)burst models 
with a Salpeter IMF (upper three curves) and a steeper IMF ($\alpha=3.3$,
lower two curves). The dotted curves show models for $Z=0.04$.
The short dashed line corresponds to an extended burst of duration $\Delta t=$ 4 Myr
(Salpeter IMF, $Z=0.02$.)
The observations are plotted at arbitrary \mup\ using the same symbols as in
Fig.\ \protect\ref{fig_std}.
The observed maximum ($\log \whb \sim$ 2.2--2.4) indicates \mup\ $\sim$ 80--90
\msun\ for a Salpeter slope, and $\protect\ga$ 120 \msun\ for $\alpha=$3.3, or somewhat
lower values for extended bursts.
}
\label{fig_imf}
\end{figure}
 
The maximum observed equivalent width of hydrogen 
recombination lines (e.g.\ \whb) of a large sample of star forming regions provide
in principal a constraint on the upper mass cut-off of the IMF (e.g\ Kennicutt
1983, Leitherer \etal\ 1999). In practice, however, as well known but not understood 
to date, there is a notable absence of regions with \whb\ as large as predicted 
for very young bursts with ages between $\sim$ 0 to 1.5--2 Myr.
As the onset of the WR phase is expected after $\sim$ 2--3 Myr quite independently
of the exact adopted stellar tracks (e.g.\ different mass loss scenarios)
this sets a new clock, and therefore the \hb\ equivalent width of the youngest 
observed WR region also contains information on the value of \mup.

In Fig.\ \ref{fig_imf} we show the dependence of the predicted \whb\ at the 
beginning of the WR phase (i.e.\ the maximum of \whb\ during this phase)
on the upper mass cut-off for different IMF slopes in instantaneous bursts.
The maximum \whb\ depends little on metallicity (see the dotted lines)
and on the choice of stellar tracks (not shown here).
Overplotted are the observed \whb\ in our WR region sample (triangles) and
the sample of BK02 (squares) drawn at arbitrary \mup.
The observed max(\whb) ($\log \whb\ \sim$ 2.2--2.4) indicates an upper
mass cut-off of $\sim$ 80--90 \msun\ for a Salpeter IMF or $\mup \ga$ 120 
\msun\ for a steeper IMF with $\alpha=$3.3.
From all the above considerations (Sect.\ \ref{s_models}) flatter slopes seem excluded.
If the bulk of the regions were forming stars in extended bursts 
the deduced value of \mup\ has to be lower; for the example illustrated here
(burst duration $\Delta t=$ 4 Myr) this would correspond to $\mup \sim$ 60 \msun\
for the Salpeter IMF.

It is important to note that the value of \mup\ derived in this way represents
a {\em lower limit}. This is the case since all observational effects known to 
affect potentially the \hb\ equivalent width (loss of photons in slit or leakage, 
dust inside \hii\ regions, differential extinction, underlying population)
can only reduce the observed \whb. The observed \whb\ represent therefore
lower limits when compared to evolutionary synthesis models.
We thus conclude that the available \whb\ measurements in metal-rich \hii\ regions
with WR stars yield a {\em lower limit of $\mup \ga$ 60--90 \msun} for the 
upper mass-cut off of the IMF. Larger values of \mup\ are not excluded.
This result is also compatible with our favoured models presented in Sect.\
\ref{s_models} (see Fig.\ \ref{fig_wnl_mod}).
Our new estimate of \mup, based only on a sample of WR regions, provides a more 
stringent limit than previous studies (SGIT00, BK02).

\section{Conclusions}
\label{s_conclude}

We have obtained high quality 
FORS1/VLT optical spectra of 85 disk \hii\ regions in the 
nearby spiral galaxies NGC 3351, NGC 3521, NGC 4254, NGC 4303, and NGC 4321.
This sample, consisting in particular of a good fraction of objects
with oxygen abundances presumably above solar (as estimated from $R_{23}$ using 
the calibration reported by Kobulnicky \etal\ 1999),
provides an unprecedented opportunity to study stellar populations,
nebular properties and ISM abundances in \hii\ regions at the high 
metallicity end.
In this first paper we have presented the observational findings on 
spectral signatures from massive stars, and compared these with
evolutionary synthesis models with the main aim of constraining
the upper part of the IMF.

The average metallicity of our \hii\ region sample is \oh\ $\sim 8.9 \pm 0.2$
using the calibration of Kobulnicky \etal\ (1999).
For 12 regions we are able to determine the electron temperature
from the transauroral \oii\ $\lambda$7325 line, yielding lower
limits on O/H (Sect.\ \ref{s_props}).
For 6 regions we have been able to confirm a high metallicity 
(\oh\ $\ga$ 8.8--8.9).
Detailed photoionisation modeling will be undertaken in the future 
to improve our abundance determinations and to include the full sample
of \hii\ regions.

The spectra of a large number (27) of regions show clear signatures
of the presence of Wolf-Rayet (WR) stars as indicated by broad emission
in the blue WR bump ($\sim$ 4680 \AA).
Including previous studies (Castellanos 2001, Bresolin \& Kennicutt 2002,
Castellanos \etal\ 2002b) 
our observations now nearly quadrupel the number of metal-rich \hii\ regions
where WR stars are known.
Approximately half (14) of the WR regions also show broad \Civ\ emission
attributed to WR stars of the WC subtype.
The simultaneous detection of \Ciii\ emission in 8 of them allows us to
determine an average late WC subtype ($\sim$ WC7-WC8) compatible
with expectations for high metallicities (Sect.\ \ref{s_props}).

Combined with existing observations of WR regions and WR galaxies at sub-solar
our data confirm the continuation of previously known trends of 
increasing WRbump/\hb\ intensity with metallicity, establish also 
such a trend for \wwr, and allow us to estimate the trend of the 
WC/WN ratio with \oh\ in extra-galactic \hii\ regions (Sect.\ \ref{s_props})

The observed strength of the blue WR bump (relative line intensities and equivalent
widths) shows quite clear trends with \whb.
Both \wwr\ and \iwr\ are found to be smaller than ``standard'' predictions from
state-of-the-art evolutionary synthesis models (Schaerer \& Vacca 1998) at 
corresponding metallicities.
Various possibilities (including deviations of the IMF from a Salpeter slope and
a ``normal'' high upper mass cut-off) which could explain this discrepancy have 
been discussed. The most likely solution is found with an improved prescription
to predict the line emission from WN stars in synthesis models 
(Sect.\ \ref{s_models}).
Using this new prescription the observed WR features are found to be broadly
consistent with short bursts and a ``standard'' Salpeter IMF extending to
high masses, as indicated by earlier studies at sub-solar metallicities.

Independently of the difficulties encountered to model the WR features
in detail, the availability of a fairly large sample of metal-rich WR regions
allows us to improve existing estimates (Schaerer \etal\ 2000, Bresolin \&
Kennicutt 2002) of the upper mass cut-off of the IMF.
Independently of the exact tracks and metallicity we derive 
a {\bf lower limit for \mup\ of 60--90 \msun} in the case of a Salpeter
slope, and larger values for steeper IMF slopes,
from the observed maximum \hb\ equivalent width of the WR regions.
This constitutes a lower limit on \mup\ as all observational effects known to 
affect potentially the \hb\ equivalent width (loss of photons in slit or leakage, 
dust inside \hii\ regions, differential extinction, underlying population)
can only reduce the observed \whb. 
From our probe of the massive star content we therefore conclude that 
there is at present no direct evidence for systematic variations of the upper mass 
cut-off of the IMF in metal-rich environments, in contrast to some claims
based on indirect nebular diagnostics (e.g.\ Goldader \etal\ 1997, 
Bresolin \etal\ 1999, Coziol \etal\ 2001).
What the origin of this ``universality'' of the IMF is, remains an open
question.

\begin{acknowledgements}
We thank Paranal staff for assistance and carrying out of the
service observations.
DS is pleased to thank Miguel Cervi\~no, Thierry Contini, Jean-Francois Le Borgne, 
and David Valls-Gabaud for various interesting discussions, and Alessandro Boselli
for comments on the Virgo structure.

\end{acknowledgements}


\end{document}